\documentclass[prl,aps,showpacs, twocolumn, floatfix,longbibliography,superscriptaddress]{revtex4-1} 

\usepackage {graphicx} 
\usepackage {amssymb} 
\usepackage {amsmath}

\newcommand{\reffig}[1]{Fig.~\ref{#1}}
\newcommand{\refeq}[1]{Eq.~(\ref{#1})}
\newcommand{\refeqs}[2]{Eqs.~(\ref{#1})-(\ref{#2})}

\newcommand{\vect}[1]{\mathrm{\mathbf{#1}}} 
\DeclareMathOperator{\imag}{Im}
\DeclareMathOperator{\tr}{tr}

\newcommand{\bra}[1]{\ensuremath{\left\langle#1\right|}}
\newcommand{\ket}[1]{\ensuremath{\left|#1\right\rangle}}
 
 \newcommand{\matrixel}[3]{\ensuremath{\left\langle #1 \vphantom{#2#3}
     \right| #2 \left| #3\vphantom{#1#2} \right\rangle}}

\begin{document}

\title{Metallic nanostructures as electronic billiards for nonlinear
terahertz photonics }

\author{Ihar Babushkin}
\affiliation{Institute of Quantum Optics, Leibniz University Hannover,
  Welfengarten 1, 30167 Hannover, Germany} 
\affiliation{Max Born Institute, Max-Born-Strasse 2a, 10117, Berlin, Germany}
\affiliation{Cluster of Excellence PhoenixD (Photonics, Optics, and
  Engineering -- Innovation across Disciplines), Welfengarten 1, 30167
  Hannover, Germany}

\author{Liping Shi} 
\affiliation{School of Engineering, Westlake University, 18 Shilongshan Road,
Hangzhou, 310024, China} 
\affiliation{Institute of Advanced Technology, Westlake Institute for Advanced Study, 18 Shilongshan Road, Hangzhou, 310024, China}

\author{Ayhan Demircan}
\affiliation{Institute of Quantum Optics, Leibniz University Hannover,
  Welfengarten 1, 30167 Hannover, Germany} 
\affiliation{Cluster of Excellence PhoenixD (Photonics, Optics, and
  Engineering -- Innovation across Disciplines), Welfengarten 1, 30167 Hannover, Germany}

\author{Uwe Morgner}
\affiliation{Institute of Quantum Optics, Leibniz University Hannover,
  Welfengarten 1, 30167 Hannover, Germany} 
\affiliation{Cluster of Excellence PhoenixD (Photonics, Optics, and
  Engineering -- Innovation across Disciplines), Welfengarten 1, 30167 Hannover, Germany}

\author{Joachim Herrmann}
\affiliation{Max Born Institute, Max-Born-Strasse 2a, 10117, Berlin, Germany}

\author{Anton Husakou}
\affiliation{Max Born Institute, Max-Born-Strasse 2a, 10117, Berlin, Germany}

\date{\today}

\begin{abstract}
  The optical properties of metallic nanoparticles are most often
  considered in terms of plasmons, the coupled states of light and
  quasi-free electrons. Confinement of electrons inside the
  nanostructure leads to another, very different type of resonances.
  We demonstrate that these confinement-induced resonances typically
  join into a single composite "super-resonance," located at
  significantly lower frequencies than the plasmonic resonance. This
  super-resonance influences the optical properties in the
  low-frequency range, in particular, producing giant nonlinearities.
  We show that such nonlinearities can be used for efficient
  down-conversion from optical to terahertz and mid-infrared
  frequencies on the sub-micrometer propagation distances in
  nanocomposites. We discuss the interaction of the
  quantum-confinement-induced super-resonance with the conventional
  plasmonic ones, as well as the unusual quantum level statistics,
  adapting here the paradigms of the quantum billiard theory and
  showing the possibility to control the resonance position and width
  using the geometry of the nanostructures.
\end{abstract}


\maketitle

\textit{Introduction.} Light propagating in the vicinity of metallic
surfaces or metallic nanostructures is strongly coupled to the
``electronic fluid'' formed by quasi-free electrons in the metal,
resulting in joint electron-photon excitations which are called
surface or particle plasmons, depending on the geometry. Plasmons, and
the corresponding plasmonic resonance (PR) are at the very heart of
optics of nanostructures.

PRs appear by matching of the incoming light to the intra-particle
fields, leading to strong surface charges and resonance peaks of
linear and nonlinear response of metallic nanoparticles at certain
frequencies \cite{mie1908,kreibig13:book,kim10c}. That is, PRs are
defined via the matching condition of the fields, rather than
electrons themselves, and have no direct relation to the electron
confinement inside the nanostructure. Plasmonic resonances are located
at quite high frequencies, commonly in the visible range.

As soon as we consider very small metallic nanoparticles, quantum
confinement of electrons in the finite volume of a nanoparticle comes
into play. Possible confined-based resonances have rarely attracted
attention \textit{per se}, separately from the properties of plasmons.
On the other hand, any calculation of the properties of small
nanoparticles does in principle include quantum mechanical confinement
of electrons as an ingredient, noticeably influencing the position and
the width of PR
\cite{kawabata66,ruppin76,kraus83,kreibig13:book,amendola17}. In the
recent years, huge progress was made in both calculations and
measurements
\cite{scholl,morton11,philip12,kreibig13:book,boyd14,qian16,amendola17,varas,zhou21}
of properties of small nanoparticles. Theoretical approaches, starting
from relatively simple analytical techniques
\cite{wood82,hache86,sato,kawabata66}, through the
single-electron-in-box \cite{wood82,genzel75}, jellium model
\cite{brack93}, hydrodynamic-like equations
\cite{ginzburg14-hydrodynamics-nanostr,hurst14-hydrodyn-nanostr} and
quantum hydrodynamic theory \cite{Takeuchi} towards direct ab-initio
DFT methods fully taking into account the ionic core structure
\cite{morton11,zhang17,zhou21,barbry,rossi} are used; for a review see
\cite{varas}. Whereas analytical and hydrodynamic approaches can
address nonlinear properties
\cite{hache86,ginzburg14-hydrodynamics-nanostr,hurst14-hydrodyn-nanostr},
complex ab-initio methods focus solely on the linear susceptibility,
unless very small nanoclusters are considered \cite{day10,day16}.

For very small nanoclusters and nanoparticles (few hundreds of atoms
and below), the response is molecular-like \cite{morton11,zhou21},
typically including a cacophony of resonances replaced by a single PR
\cite{zhou16,zhou21,townsend}. At lower frequencies, a prominent
molecular-like resonance is the HOMO-LUMO transition
\cite{barcaro06,kwak17,zhou19,zhou21}. It describes an excitation of a
single localized electron, and depends heavily on the molecular
structure \cite{barcaro06}.

As we shift to larger nanoparticles, the HOMO-LUMO transition
effectively fades out \cite{zhou21} because its decay rate grows
exponentially with the size of the nanostructure \cite{kwak17,zhou19}.
However, the PR is not the only one which remains. Signatures of a
single resonance well below the plasmonic one but above the HOMO-LUMO
transition were observed experimentally \cite{oates05,wyrwas07} and
later confirmed theoretically \cite{scholl,he10} involving ab-initio
simulations \cite{he10}. This resonance is however  mostly overlooked since then,
and there is no consensus in explanations of its nature. Whereas in
\cite{scholl,wyrwas07} the confinement-based argument were put
forward, \cite{oates05,he10} tries to explain it without leaving the
plasmonic framework by introducing "restoring force on the electrons".

Here we show that confinement-based resonances under normal conditions
(at least for nanostructures with regular enough geometry) merge
together and form a single broadband ``super-resonance'' located
commonly in THz and MIR frequency range, with both the width and
position widely controllable with the nanostructure geometry. Such
super-resonance must be considered as one of the universal signatures
of metallic nanostructures, yet fundamentally different from the PR.

We study the statistical features of confinement-based resonances by
adapting and modifying the paradigm of neighboring-level statistics
from the field of quantum billiards \cite{stoeckmann99:book}.
Electrons confined in a nanostructure certainly represents a type of
quantum billiard. Yet, up to now, with only very few exceptions
\cite{simons93,ravnik21}, non-metallic billiards such as semiconductor
quantum dots
\cite{nakamura88,jalabert90-chaos-quant-dots,akis97,zozoulenko97,burke10-scars-quantum-dots-experiment,ponomarenko08,kucharik19}
were considered. As we show, the metallic nature of our billiard
provides an unique opportunity to observe certain features the level
statistics directly in the optical properties. 

Furthermore, the super-resonance provides a broadband nonlinear
response, leading to giant nonlinearities. We demonstrate how these
nonlinearities can be used for an efficient optical rectification and
difference-frequency mixing in the nanocomposites, enabling broadband
conversion from optical to MIR or THz ranges by sub-micrometer
devices.

\begin{figure}[t!] 
\includegraphics[width=0.99\columnwidth]{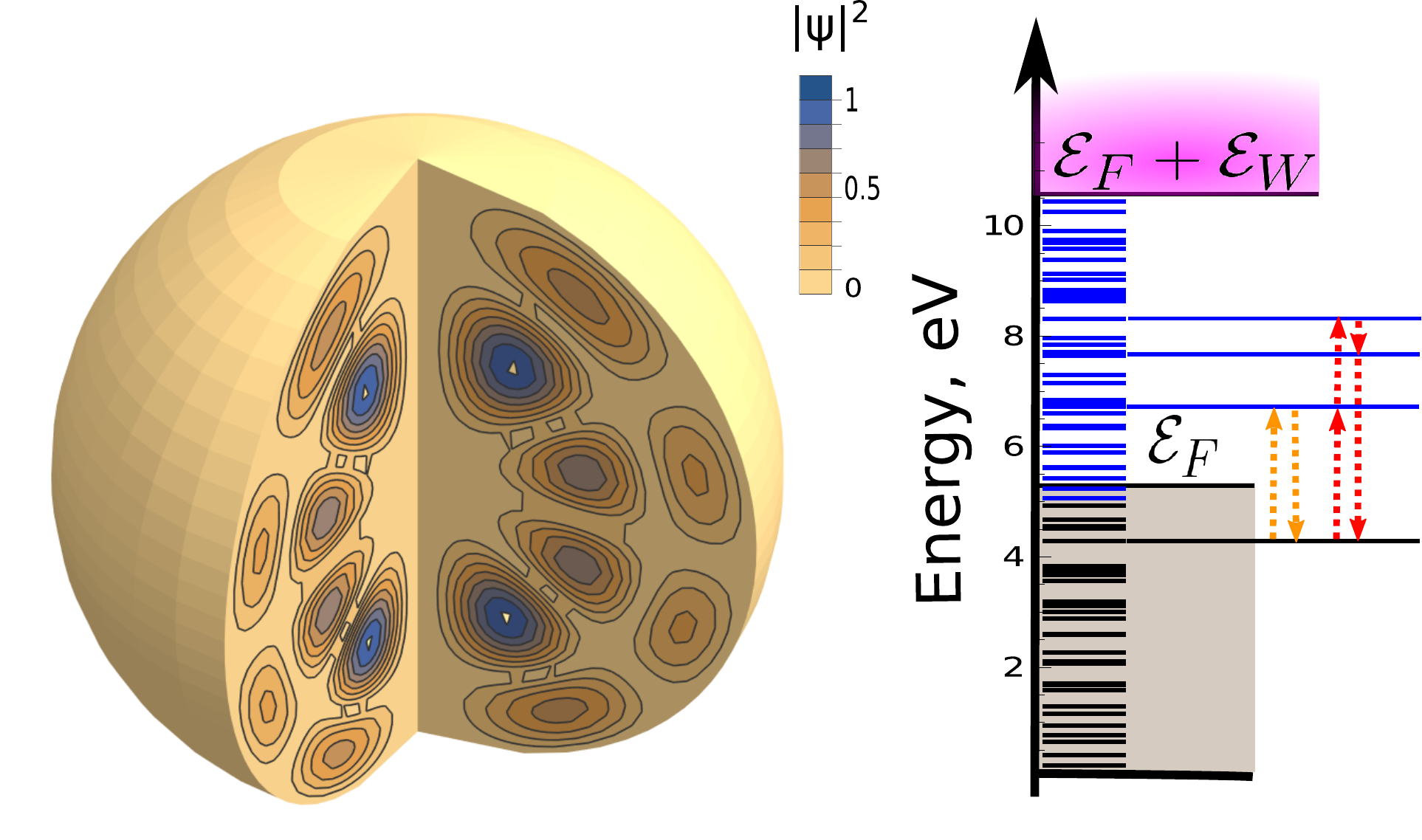}
\caption{ \label{fig:base} A spherical gold nanostructure
  (left) of 2.75 nm diameter and the band level structure (right)
  appearing due to the electron confinement (quantum billiard
  resonances). Fermi energy $\mathcal E_F$ and work function
  $\mathcal E_W$ as well as exemplary virtual transitions contributing
  to linear (orange arrows) and nonlinear (red arrows) properties are
  indicated. One of the wavefunctions $\psi(x,y,z)$ is visualized
  inside the nanostructure (a sector of the sphere is cut for
  visualization purposes). Even for the simplest nanoparticles, energy
  level structure is quite complex. }
\end{figure}

\textit{The model.} We used a simple analytical approach of a
single-particle-in-a-box \cite{kraus83, hache86,wyrwas07,scholl} (see
also more details in Supplementary) with electron fully confined
inside a nanoparticle. For calculation of optical properties, we
consider electron energy structure characterized by Fermi energy
$\mathcal E_F$ and work function of $\mathcal E_W$, as illustrated in
Fig. 1 right. This approximation works well for metals with a simple
Fermi surface such as alkali metals (Li, Na, Ca, Rb), it is an
acceptable simplification for metals with somewhat more complicated
Fermi surfaces such as Cs, Cu, Ag or Au, and is barely applicable at
all for other metals. The linear ($\chi^{(1)}$) and nonlinear
($\chi^{(3)}$) susceptibilities were calculated using a version of the
standard perturbative iterative approach \cite{boyd08:book} which
takes into account selection rules following from the Fermi-Dirac
statistics (see Supplementary). The linear and nonlinear optical
properties can be described via a sum of contributions of virtual
transitions from inside the Fermi sea to outside and back [see orange
(linear) and red (nonlinear) arrows \reffig{fig:base}]. We also
assumed fast population decay time $T_1=50$ fs and dephasing time
$T_2=5$ fs \cite{hache86}.

To be more specific, in our numerical simulations we consider gold
since it is a very widespread material, and is suitable for composites
due to low imaginary part of susceptibility. Yet we note that
conclusions we draw below are basically metal-independent (taking into
account precautions mentioned above). In gold, a simple ideal-metal
picture discussed above neglects several linear and nonlinear effects,
such as interband transitions, influence of finite temperature, and
hot-electron nonlinearities. However, these effects are negligible for
low-frequency response in THz or MIR range driven by femtosecond
pulses, and our model remains adequate in this regime (see
Supplementary for justifications and detailed estimates).

\begin{figure*}[thb]
\includegraphics[width=\textwidth]{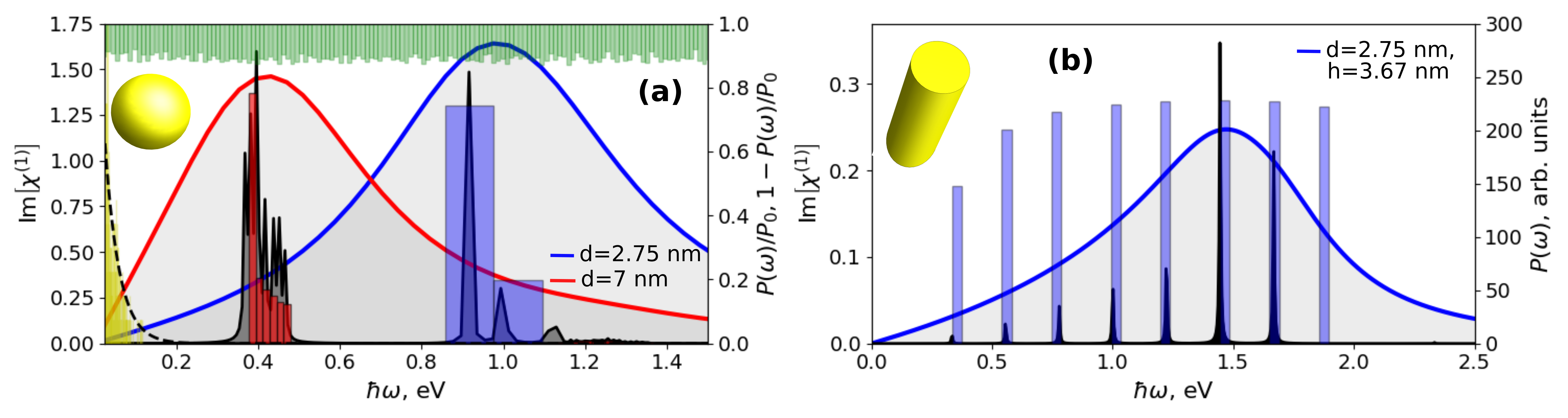}
\caption{ \label{fig:lin} Billiard resonances and weighted level
  statistics for a golden sphere of few exemplary diameters $d$ (a)
  and a cylinder (b) of the diameter $d$ and height $h$. The blue and
  red curves in (a) and (b) show $\imag{(\chi^{(1)})}$ as the function
  of frequency for different sizes (see legend). Bars of the same
  colors show the corresponding weighted level statistics $P(\omega)$
  normalized to some $P_0$ ($P_0$ is not the same for different
  curves). Green bars show the statistics for all allowed transitions
  attached, for visibility, to the upper $x$-axis (that is,
  technically, $1-P(\omega)/P_0$ is shown; see also larger frequency scales
  in Supplementary). Black lines show $\imag{(\chi^{(1)})}$ for $T_2$
  increased 100 times. Yellow bars show NLS and dashed black line
  indicates the Poissonian statistics. }
\end{figure*}

\textit{Quantum billiard (confinement-based) resonances.} The typical
level structure obtained by the above model for an exemplary spherical
gold nanoparticle of the diameter $d=2.75$ nm is shown in
\reffig{fig:base}. These levels originate from electron confinement in
the nanostructure. Even in the presented case of a very simple
particle, the levels look quite irregular. This is a familiar picture
in the framework of of quantum billiard theory
\cite{stoeckmann99:book}, were the statistical properties of the level
distribution play one of the central roles. For instance, one can
consider the neighboring level statistics (NLS), which allows to
distinguish between integrable (regular) and non-integrable (chaotic)
billiards. For the regular billiards, such as spheres, the probability
density $P( \omega)$ of neighboring-level distance $ \omega$ obeys
Poissonian statistics: $\ln{P( \omega)}\propto - \omega$. This is also
true in our case: NLS corresponding to \reffig{fig:base} is plotted in
\reffig{fig:lin}(a) by yellow bars and coincides well with the
Poissonian distribution (black dashed line).

Judging from such statistics, one might expect a conglomerate of
resonances near zero frequency, but this is not the case. An optical
response $\chi^{(1)}$ resulting from the electron confinement for few
exemplary nanostructures is shown in \reffig{fig:lin}. Note that, in
addition to the confinement-based impact shown in \reffig{fig:lin},
the full linear response includes also the so-called Drude part,
representing the action of quasi-free-electrons (see Supplementary for
more details).

The clearly observed feature of the confinement-based linear response
is the presence of a single resonance-like peak in the IR/THz range at
a non-zero resonance frequency $\omega_{\mathrm{conf}}$ which quickly
decreases with increasing particle size. Such peak was observed
experimentally \cite{oates05,wyrwas07} and theoretically
\cite{scholl,he10}. It is easy to see that this resonance coincides
well with the minimally possible allowed transitions close to the
Fermi energy which can be, for a spherical nanoparticle, analytically
estimated as (see Supplementary for details), cf. also
\cite{wyrwas07}):
\begin{equation}
  \omega_{\mathrm{conf}} \approx \frac{\pi}{r}\sqrt{\frac{\mathcal E_F}{2m_e}},\label{eq:omega_conf}
\end{equation}
where $m_e$ is the electron mass. This makes it fundamentally
different from the PR, being located at a significantly lower
frequency. Yet, why does only a single resonance arise? Can we
influence it width and position? These and related questions will be
addressed in the following paragraphs.

Closer consideration allows to establish that, starting already from
quite small nanosphere diameters $d\approx 2.5$ nm, the resonance near
$\omega_{\mathrm{conf}}$ is composed of many transitions with nearby
frequencies [see \reffig{fig:lin}(a)].
These
transitions merge into one single ``super-resonance''.
This is especially well observable if we consider much larger $T_2$
(which would correspond to low temperatures \cite{kabanov08}). In this
case, many separated resonances are indeed visible in the optical
response, as shown in Fig. 2(a,b) by black curves.
The particular structure of the transitions depends significantly on
the geometry [cf. \reffig{fig:lin}(b) for a cylinder]. For instance,
the position and width of the super-resonance for a cylinder are
shifted in comparison to a sphere with the same volume and diameter by
the noticeable amount of 35\% and 23\%, correspondingly. Nevertheless,
Eq.~1 remains a valid, yet rough estimation of the position of the
resonance.

Both the position of the super-resonance $\omega_{\mathrm{conf}}$ and
its structure can be analyzed using the level-distance statistics
similar to NLS. A na\"ive approach would be to calculate such
statistics using all dipole-allowed transitions between the
confinement-based levels, shown in downward green rectangles in Fig.
2(a), and covering extremely broad range around 10 eV (see green bars
in Fig. 2(a) and also Supplementary). However, it must be modified to
include only transitions from below to above of $\mathcal E_F$,
obviously corresponding to the Pauli principle and absence of
population above the Fermi level. In addition, in the statistics we
weight the transitions by the square of the corresponding dipole
momentum, thus taking into account the known tendency of the
transition dipole momenta to rapidly decrease, on average, with the
energy difference. The resulting modified statistics is shown by the
red and blue bars in \reffig{fig:lin} and agrees nicely (for large
$T_2$) both with the position of the super-resonance and with its
width for small $T_2$. The super-resonance has a certain ``natural''
width, for the case of nanospheres it can be estimated as
$\omega_{\mathrm{conf}}/4$ (see Supplementary). 
Analysis of the position 
of the super-resonance
for nanospheres (see Supplementary) indicates that the energy of the
participating states is located mostly in radial (rather than angular)
motion.

\begin{figure}[thb]    
\includegraphics[width=\columnwidth]{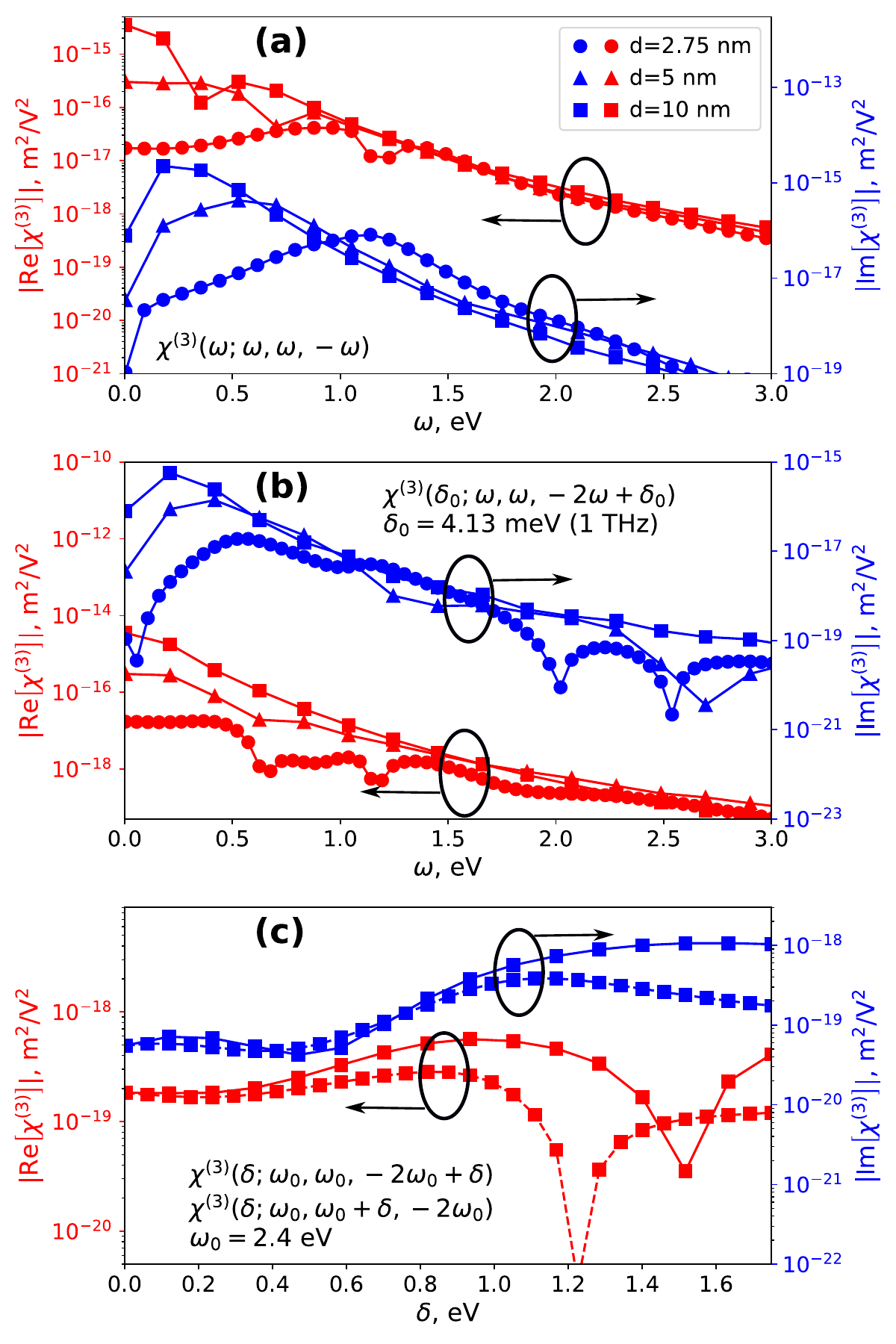}
\caption{ \label{fig:chi3} Nonlinear susceptibility for the Kerr
  nonlinearity (a) and for different FWM processes leading to the
  optical rectification (b,c) in dependence on the signal $\delta$
  (c) and pump $\omega$ (b) frequency, shown for the nanostructures of
  different diameters.  }
\end{figure}

\textit{Nonlinearities.} The above described low-frequency resonance
is expected to lead also to strong nonlinearities; in our case,
$\chi^{(3)}\ne 0$ as calculated using the approach described above. We
note that such approach to calculate Kerr nonlinearity was already
utilized in \cite{hache86}, however, instead of discrete spectrum,
approximation of continuous density of states was used. Nevertheless,
we checked that our calculations are in quantitative agreement with
\cite{hache86}; they are also in agreement with experimental
measurements for short pulses (see \cite{boyd14} and references
therein). An example of $\chi^{(3)}(\omega;\omega,\omega,-\omega)$ for
the four-wave-mixing (FWM) process $\omega + \omega -\omega = \omega$
(corresponding to the Kerr nonlinearity) is shown in
\reffig{fig:chi3}(a) for several diameters. The low-frequency resonance we
observed in $\chi^{(1)}$ is also well-visible here. Whereas in the
linear response the Drude part dominates (see Supplementary), in the
nonlinear response it is fully absent.

We now try to exploit this low-frequency resonance. Motivated by
detection and spectroscopic applications of THz and MIR radiation, we
focus on the FWM providing a signal in THz and MIR range, generated
from a sub-100 fs pump pulse. Nonlinear susceptibilities
$\chi^{(3)}(\delta;\omega,\omega,-2\omega+\delta)$ and
$\chi^{(3)}(\delta;\omega,\omega+\delta,-2\omega)$, leading to
generation of low-frequency signal at frequency $\delta$ as a result
of a FWM process in a two-color pump at frequencies $\omega$
(fundamental) and $2\omega$ (its second harmonics) are presented in
\reffig{fig:chi3}(b-c). Both Kerr rectification nonlinearities
presented in \reffig{fig:chi3} are several orders of magnitude higher
than the Kerr nonlinearity of the fused silica $\sim 2\times 10^{-22}$
m$^2$/V$^2$. In \reffig{fig:chi3}(a), where pump-frequency
dependencies are shown, the billiard super-resonance described above
is very clearly visible. This is not a unique property of metallic
nanostructures. Giant nonlinearities in semiconductor nanostructures
due to billiard resonances were recently predicted in
\cite{Kucharik21}.

\begin{figure*}[thb]
\includegraphics[width=\textwidth]{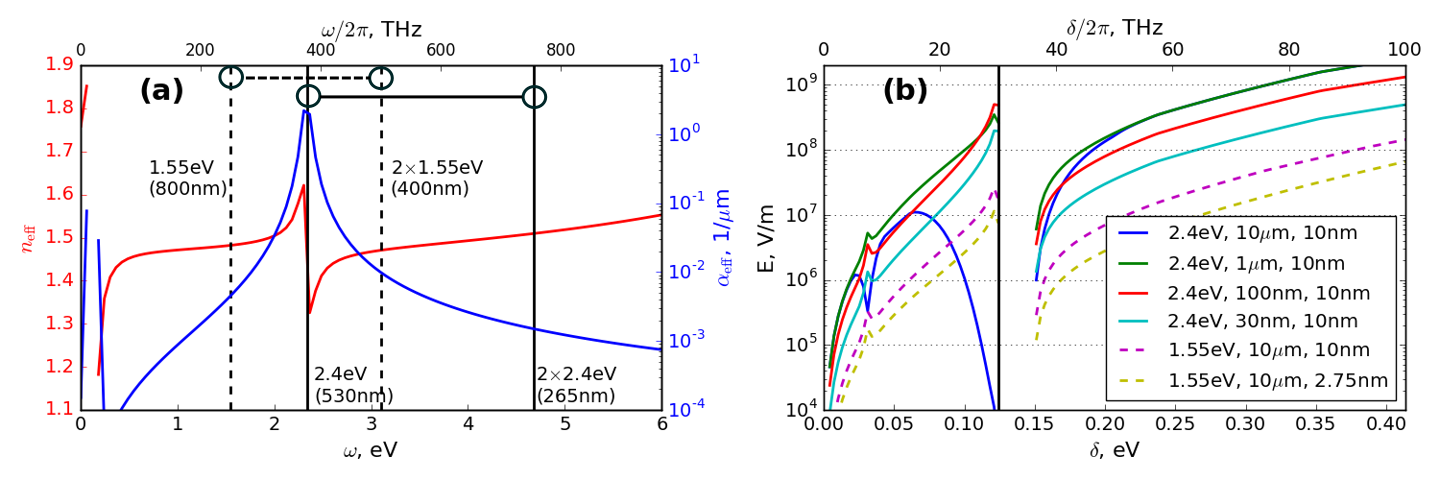}
\caption{ \label{fig:prop} Efficient generation of THz and MIR light
  in a composite of gold spheres with 2-color pump. (a) effective
  refractive index $n_\mathrm{eff}$ and effective losses
  $\alpha_\mathrm{eff}$ in dependence on frequency $\omega$ for a gold
  nanostructures with $r=10$ nm immersed into fused silica ($f=0.01$).
  Vertical lines show two variants of the 2-color pump at
  $\omega_0= 1.55$ eV ($\lambda_0=530$ nm, dashed line) and
  $\omega_0= 2.4$ eV ($\lambda_0=530$ nm, solid line). The horizontal
  lines connect the spectral components of two-color pump. (b) The
  generated field amplitude for different propagation distances $L$
  and nanostructure diameter $d$ (see legend) and the pump as described
  in text. Solid vertical line in (b) separates THz from MIR band. }
\end{figure*}

\textit{Efficient frequency difference generation.} As an interesting
application we consider the process of optical rectification and
difference frequency generation, governed by three nonlinearities
$\chi^{(3)}(\delta;\omega_0,\omega_0,-2\omega_0+\delta)$,
$\chi^{(3)}(\delta;\omega_0+\delta,\omega_0,-2\omega_0)$ and
$\chi^{(3)}(\delta;\omega_0,\omega_0+\delta,-2\omega_0)$, with a
two-color optical pump at around $\omega_0$ and $2\omega_0$ and signal
$\delta \ll \omega_0$ in THz and MIR.
We solve the propagation equations, assuming slowly varying envelope
approximation and taking into account dispersion relations, but
neglecting nonlinear effects for the pump waves because of very very
small propagation distance (see Supplementary for details). 
Both $\chi^{(3)}_{\mathrm{eff}}$ and the linear susceptibility
$\chi^{(1)}_{\mathrm{eff}}$ are calculated from given linear and
nonlinear properties of the nanoparticles ($\chi^{(1)}_\mathrm{NP}$,
$\chi^{(3)}_\mathrm{NP}$) and host ($\chi^{(1)}_\mathrm{h}$,
$\chi^{(3)}_\mathrm{h}$) using the effective medium approach
\cite{zeng88-effective-n-n2-nanoparticles} (see Supplementary). By
calculation of the linear properties the full linear susceptibility
containing both confinement-based and Drude parts are included. As a
host material, we take fused silica which possesses strong losses in
the range between 30 and 40 THz (see \reffig{fig:prop}), but otherwise
is quite transparent \cite{palik98:book}. We consider the filling
factor of $f=0.01$ and neglect the nonlinearity of the host. Resulting
effective linear quantities are shown in \reffig{fig:prop} and
demonstrate the usual PR resonance at around 2.4 eV with the width of
around 30 THz. The shortest pulses still supported by this resonance
are around 30 fs in duration.

Assuming an exemplary pulse durations of around 30 fs, we must consider two
regions for the pump where conversion works significantly different.
For the signal in the THz range ($\delta/2\pi \le$ 30 THz), the
frequencies $j\omega_0$ and $j\omega_0+\delta$ ($j=1,2$) are both
located within the spectrum of the pump. In contrast, for the signal
in MIR range $\delta/2\pi >$ 30 THz, the components $\omega_0+\delta$,
$2\omega_0+\delta$ are not within the pump spectrum anymore. This
leads to different treatment of these two frequency ranges for the
selected pulse duration (see Supplementary). 

The resulting field amplitude at 0th harmonic $A_0$ is given in
\reffig{fig:prop}(b) for different parameters and for the pump
amplitudes $A_1=A_2=10^{10}$~V/m. This pump for 30-fs pulses
corresponds to a fluence around 0.3 J/cm$^2$, which is yet below the
damage threshold of gold (around 0.5 J/cm$^2$
\cite{poole13-damage-gold}) and of fused silica (around 1 J/cm$^2$
\cite{chimier11-damage-silica}).
One can see that in THz range the signal amplitude reaches
$5\times 10^{8}$ V/m corresponding to efficiency of around $10^{-3}$.
In MIR range, the amplitude can exceed $10^{9}$ V/m, delivering
efficiencies above the percent level. Moreover, the maximal efficiency
is achieved at 100 nm propagation distance for THz signal and 1 $\mu$m
for MIR signal. From \reffig{fig:prop}(b) one can also see that the
most efficient conversion is achieved for the pump frequency
$\omega_0$ centered at the PR (solid lines in \reffig{fig:prop}). In
this case, the coupling of the pump to the signal is most efficient.

\textit{Discussion and conclusions.} We showed that confinement-based
energy levels in metallic nanostructures, representing an integrable
(or close to integrable) quantum billiards, typically join together
into a single "super-resonance", which position and width can be
controlled by the geometry of the nanostructure. Whereas we focused
here on (almost) integrable quantum metallic billiards, we anticipate
richer resonance structure and control possibilities if truly chaotic
billiards are considered. We analyzed the super-resonance, using the
level statistics extended in comparison to typically used in quantum
chaos theory. 
In the linear regime the ballistic super-resonances is "hidden" behind
the much stronger Drude response, yet it manifests itself strongly in
a giant nonlinearity. This nonlinearity can be in addition enhanced by
interaction with plasmons and effectively used to down-convert light
to THz and MIR range with high efficiency already after 100-nm
distances, despite of huge linear and nonlinear losses. Our
confinement-based framework might also be helpful in a deeper
understanding of the recent experimental work on efficient THz
generation in nanostructures \cite{luo14,keren-zur19} and paves a way
to extend newly proposed electronic meta-devices \cite{samizadeh23}
into the nonlinear regime.

\begin{acknowledgments}
IB, AD and UM acknowledge support from the Deutsche Forschungsgemeinschaft  (DFG) under
Germany's Excellence Strategy within the Cluster of Excellence PhoenixD
(Photonics, Optics, and Engineering – Innovation Across Disciplines) (EXC 2122, projectID 390833453). AH acknowledges support from European Union project H2020-MSCA-RISE-2018-823897 "Atlantic".
\end{acknowledgments}

\section{Supplementary}
\label{sec:supplement}

\subsection{Simplified Hamiltonian and wavefunctions for the case of
  spherical particles}

To approach the problem analytically, we consider a spherical metallic
particle of the radius $a$ (diameter $d=2a$). Since we are interested
in low frequencies, we neglect the interband transitions. We neglect
the temperature effects assuming that $N$ electrons in conduction band
occupy all levels below Fermi energy $\mathcal E_F$ (see Fig. 1 of the
main article). The energy structure is calculated assuming
one-electron approximation and corresponds to that of a single free
electron in a infinite-strength spherical potential of the radius $a$.
To take into account finiteness of the potential, the levels above the
work function $\mathcal E_W$ are disregarded. The validity of these
approximations is justified below.

The corresponding single-particle eigenproblem can be then formulated
as $\hat H_0\psi = \mathcal E \psi$, with
$H_0 = -\hbar^2/2m_e \Delta + V(r)$, $V(r)= 0$ for $r\le a$ and
$V(r) = \infty$ for $r>a$ ($m_e$ is the electron mass). With the
approximations above, we neglect various effects of electron-electron
and electron-ion interaction such as interband transitions, electron
heating, the change of the eigenstates due to the finite height of the
potential, and other effects, which play only minor role at low
frequencies and ultrashort sub-100-fs pulse durations. 
The validity of this approximation is discussed in
Sec.~\ref{sec:other-mechs} below. As it will be shown there, our
simple model is rather adequate for the parameters we consider,
despite of its simplicity. The advantage of this approach is the
possibility to determine the energy structure analytically. The
corresponding eigenfunctions are combinations of spherical harmonics.
The energies are defined as
\begin{equation}
\mathcal E_{nl} = \mathcal E_0 \alpha_{nl}^2\label{eq:E}
\end{equation}
where $n,l$ are quantum
numbers,
\begin{equation}
\mathcal E_0= \dfrac{\hbar^2}{2m_ea^2},\label{eq:E0}
\end{equation}
and $\alpha_{nl}$ is
the $n$th zero of the Bessel function of order $l$. In contrast to the
Coulomb potential, there is no degeneracy in $l$.

The eigenfunctions of the problem described in the
main article are:
\begin{equation}
  \label{eq:1}
  \psi_{nlm}(r,\theta,\phi)= R_{nl}(r) Y_l^m(\theta,\phi),
\end{equation}
where $n,m,l$ are quantum numbers, $R_{nl}(r) = 
  \dfrac{\sqrt{2} j_l(\alpha_{nl}r/a)}{\sqrt{a^3}j_{l+1}(\alpha_{nl})}$, $Y_l^m$ are spherical
  functions ($-l \le m \le l$), $j_l$ is the spherical Bessel
  function of order $l$, and $\alpha_{nl}$ is its $l$th zero.

 The radial part of the matrix element
is, for the allowed transitions $l'-l=\pm 1$,
\begin{equation}
  \label{eq:3}
  \mu_{nl,n'l'} = \frac{4ae \mathcal E_0 \sqrt{\mathcal E_{nl} \mathcal
    E_{n'l'}}}{(\mathcal E_{n'l'} - \mathcal E_{nl})^2},
\end{equation}
and zero in other cases (here $e$ is the electron charge).

\subsection{Cylindrical geometry}
\label{sec:cylindrical-geometry}

In this section we determine the eigenvalues for the cylindrical geometry.
We assume here that the main axis of the cylinder oriented along
$z$-direction, and the light is assumed to be linearly polarized also
in $z$-direction. Note the difference in the denotations with the part
where propagation is considered: there, $z$-direction is the direction
of the light propagation. 

The eigenfunctions in $(r,\theta,z)$-coordinates are \cite{okamoto21:book}:
\begin{equation}
  \label{eq:cylinder-psi}
  \psi_{nlm}(r,\theta,z) = C_{nlm}J_m(r \tilde \alpha_{lm}/a) \cos(m\theta) \cos(\pi n z/h),
\end{equation}
where $C_{nlm}$ is the normalization factor, $h$ is the height of the
cylinder, $\tilde \alpha_{lm}$ is $l$th zero of the Bessel
function $J_m(x)$ of order $m$. These eigenfunctions are described by
three integer quantum numbers: $n$ describes the localization in
$z$ direction, $m$ and $l$ -- along the orthogonal directions. 

The energies of these eigenstates are given by the expression
\begin{equation}
  \label{eq:en-cyl}
  \mathcal E_{nlm} = \frac{\hbar^2}{2 m_e } \left( \frac{ \tilde
      \alpha_{lm}^2}{a^2} + \frac{4 \pi^2n^2 }{h^2} \right). 
\end{equation}
Because the light is assumed to be linearly polarized in
$z$-direction, only $z$-components of the dipole moments
\begin{equation}
  \label{eq:dm-cyl-def}
\mu_{nlm,n'l'm'} =   \bra{\psi_{nlm}}z\ket{\psi_{n'l'm'}}
\end{equation}
play a role, and they can be calculated as:
\begin{equation}
  \label{eq:mu_cyl}
  \mu_{nlm,n'l'm'} \propto \frac{2nn'\left((-1)^{n+n'}-1\right)}{(n-n')^2(n+n')^2} \delta_{ll'}\delta_{mm'}, 
\end{equation}
where we omitted for simplicity a pre-factor which comes from the
normalization of the wavefunctions.

\subsection{Derivation of Eq. 1 of the main article}
\label{sec:derivation-eq.-1}

As we see in the main article (see also below), the main role in the
super-resonance is played by the transitions close to the Fermi energy
$\mathcal E_F$. For not very small nanostructures, this corresponds to
relatively large $l$ and $n$. For large $n$ and $l$ an analytical
estimation
\begin{equation}
\alpha_{nl} \approx (2n+l)\pi/2\label{eq:alpha_nl_approx}
\end{equation}
is possible. Based on this, the energy difference between the
allowed transitions $\Delta l=\pm 1$ near the energy $\mathcal E$ is
\begin{equation}
  \label{eq:DE}
  \Delta \mathcal E = \pi\sqrt{\mathcal E \mathcal E_0}. 
\end{equation}
Near the Fermi energy $\mathcal E\approx \mathcal E_F$, substituting
\refeq{eq:E0}, we obtain Eq.~1 of the main article, with  $\hbar
\omega_{\mathrm{conf}}$ identical to $\Delta \mathcal E$.

We note that \refeq{eq:alpha_nl_approx} is valid for $n\gg l$. In
contrast, for $n\approx 1$ we have $\alpha_{nl} \approx (2n+l)$, that
is, the positions of the resonances would be shifted in this latter
case by the factor $\sim \pi/2$ to the lower frequencies. The
resonances shown in Fig. 2 of the main article as well as in
\reffig{fig:lin} coincide well with \refeq{eq:alpha_nl_approx},
indicating, that large $n$ are involved (see also
Sec.~\ref{sec:level-stat-deta}).

In addition to the derivation of the position of the super-resonance,
we are able to approximately deduce its ``natural'' shape and width.
For this, see Sec.~\ref{sec:level-stat-deta} below.

\subsection{Linear and nonlinear susceptibilities of a single
  nanostructure}

\label{sec:lin-nonlin-prop}

The corresponding expressions for $\chi^{(1)}$ and $\chi^{(3)}$ are
obtained by the method of iterations \cite{boyd08:book}: The evolution
of the density matrix $\rho$ in the presence of damping can be, under
suitable approximations, written as
\begin{equation}
  \label{eq:rho_gen}
  \dot \rho = -\frac{i}{\hbar} \left[\hat H, \rho \right] - \Gamma(\rho - \rho^{(eq)}),
\end{equation}
where $\hat H$ is the full Hamiltonian and $\rho^{(eq)}$ is the
equilibrium value for $\rho$, $\Gamma$ describes the decay. The
Hamiltonian consists of the action of the potential well $H_0$ (see
the main article) as well as the action of the field
$V = e\vect r \vect E$ (in dipole approximation).
In terms of the eigenfrequencies of $\hat H_0$, $\omega_{mn} =
(\mathcal E_m -\mathcal E_n)/\hbar$, \refeq{eq:rho_gen} can be
rewritten in terms of perturbations as:
\begin{equation}
  \label{eq:rho_perturb}
  \dot \rho_{mn} = -i\omega_{mn}\rho_{mn} - \frac{i}{\hbar}
  \left[V,\rho\right] - \gamma_{mn}(\rho_{mn} - \rho_{mn}^{(eq)}),
\end{equation}
where $\rho_{mn} = \matrixel{m}{\rho}{n}$,
$\rho^{(eq)}_{mn} = \matrixel{m}{\rho^{(eq)}}{n}$,
$\gamma_{mn} = \matrixel{m}{\Gamma}{n}$, $\ket{m}$, $\ket{n}$ are
eigenstates of $\hat H_0$ corresponding to eigenvalues $\mathcal E_m$,
$\mathcal E_n$ (note that here, in contrast to previous sections, we
denote by $n$ and $m$ the ``multiindices'', fully describing the
eigenstate). One can obtain $\rho$ iteratively, in the form of
$\rho = \rho^{(0)} + \theta \rho^{(1)} + \theta^2\rho^{(2)} +\ldots$,
where $\theta$ is a formal small parameter, assuming thereby that $V$
is a small perturbation ``of order of $\theta$''. As an initial
approximation we obtain
\begin{equation}
\rho^{(0)} = \rho^{(eq)},\label{eq:rho0}
\end{equation}
and $\rho^{(n)}$ is related to 
$\rho^{(n-1)}$ as:
\begin{equation}
  \label{eq:rho_perturb_n}
  \dot \rho^{(n)}_{mn} = -(i\omega_{mn}+\gamma_{mn})\rho^{(n)}_{mn} - \frac{i}{\hbar}
  \left[V,\rho^{(n-1)}\right].
\end{equation}

The nonlinear polarization of $n^{\rm th}$ order is defined via
$P^{(n)}_{i} =
\epsilon_0\sum_{j,k,\ldots}\chi^{(n)}_{i;jk\ldots}E_jE_k\ldots$, where
$P^{(n)}_i$, $E_i$ are the components of the vectors $\vect P^{(n)}$,
$\vect E$, respectively. $\vect P^{(n)}$ is given in terms of $\rho^{(n)}$ as
$\vect P^{(n)} = -eN\tr{\left(\rho^{(n)} \vect r\right)}$, where $N$
is concentration of the particles. The expression for $\chi^{(n)}$ is
obtained by comparing the two expressions for $\vect P^{(n)}$ above.
For the first order susceptibility $\chi_{ij}^{(1)}(\omega_p)$ we thus
obtain:
\begin{widetext}
  \begin{equation}
    \label{eq:chi1}
\chi_{ij}^{(1)}(\omega_p)= \chi^{(1)}_D + \frac{N}{\epsilon_0\hbar}\sum_n\left[\frac{\mu^i_{an}\mu^j_{na}}{(\omega_{na}-\omega_p)-i\gamma_{na}}+\frac{\mu^i_{an}\mu^j_{na}}{(\omega_{na}+\omega_p)+i\gamma_{na}} \right ].
\end{equation}
\end{widetext}
We note once more that $a$ and $n$ are ``multiindices'', that is, every of them
describes a particular set of quantum numbers $n,l,m$ fully
characterizing the system; $\omega_{mn} = \mathcal E_{mn}/\hbar$,
$\gamma_{mn}=\delta_{mn}\gamma$ ($\delta_{mn}$ is a Kronecker symbol),
$\gamma=1/T_2$; $T_2$ and $T_1$ and are given in the main article. For
the isotropic case we consider here, the indices $i,j$ in
\refeq{eq:chi1} are disregarded. According to \refeq{eq:rho0},
$\rho^{(0)}_{ll}$ describes the unperturbed populations (see more
details below). The first term in \refeq{eq:chi1} describes the Drude
dispersion. It must be introduced into \refeq{eq:chi1} as an
additional phenomenological term since its proper first-principle
treatment is possible only if electron-phonon interactions
\cite{kitamura15} are taken into account, which is not the case in our
approach.
\begin{equation}
  \label{eq:drude}
  \chi^{(1)}_D  = \chi_\infty  - \frac{\omega^2_{pl}}{\omega
    \left(\omega + i\gamma_D\right)},
\end{equation}
with $\omega_{pl} = Ne^2/(\epsilon_0 m_e)$ and the effective
phenomenological quantities are taken as $\chi_\infty = 8.84$,
$\gamma_D=0.067$ eV, and $N=5.9\times 10^{28}$ m$^{-3}$ as given in 
\cite{sonnichsen2001plasmons:phd}.
 
For the third-order susceptibility we have:
\begin{widetext}
  \begin{equation}
        \label{eq:chi3}
\begin{split}
&\chi_{kjih}^{(3)}(\omega_p+\omega_q+\omega_r;\omega_r,\omega_q,\omega_p)=\frac{N}{\epsilon_0\hbar^3}\mathcal
P_I\sum_{\nu nml}\rho_{ll}^{(0)}\\&
\times\left\{\frac{\mu_{l\nu}^k\mu_{\nu n}^j\mu_{nm}^i\mu_{ml}^h}{
[(\omega_{\nu l}-\omega_p-\omega_q-\omega_r)-i\gamma_{\nu l}]
[(\omega_{nl}-\omega_p-\omega_q)-i\gamma_{nl}]
[(\omega_{ml}-\omega_p)-i\gamma_{ml}]
}\right.\\&
+\frac{\mu_{l\nu}^h\mu_{\nu n}^k\mu_{nm}^j\mu_{ml}^i}{
[(\omega_{n\nu}-\omega_p-\omega_q-\omega_r)-i\gamma_{n\nu}]
[(\omega_{m\nu}-\omega_p-\omega_q)-i\gamma_{m\nu}]
[(\omega_{\nu l}+\omega_p)+i\gamma_{\nu l}]
}\\&
+\frac{\mu_{l\nu}^i\mu_{\nu n}^k\mu_{nm}^j\mu_{ml}^h}{
[(\omega_{n\nu}-\omega_p-\omega_q-\omega_r)-i\gamma_{n\nu}]
[(\omega_{\nu m}+\omega_p+\omega_q)+i\gamma_{\nu m}]
[(\omega_{ml}-\omega_p)-i\gamma_{ml}]
}\\&
+\frac{\mu_{l\nu}^h\mu_{\nu n}^i\mu_{nm}^k\mu_{ml}^j}{
[(\omega_{mn}-\omega_p-\omega_q-\omega_r)-i\gamma_{mn}]
[(\omega_{nl}+\omega_p+\omega_q)+i\gamma_{nl}]
[(\omega_{\nu l}+\omega_p)+i\gamma_{\nu l}]
}\\&
+\frac{\mu_{l\nu}^j\mu_{\nu n}^k\mu_{nm}^i\mu_{ml}^h}{
[(\omega_{\nu n}+\omega_p+\omega_q+\omega_r)+i\gamma_{\nu n}]
[(\omega_{nl}-\omega_p-\omega_q)-i\gamma_{nl}]
[(\omega_{ml}-\omega_p)-i\gamma_{ml}]
}\\&
+\frac{\mu_{l\nu}^h\mu_{\nu n}^j\mu_{nm}^k\mu_{ml}^i}{
[(\omega_{nm}+\omega_p+\omega_q+\omega_r)+i\gamma_{nm}]
[(\omega_{m\nu}-\omega_p-\omega_q)-i\gamma_{m\nu}]
[(\omega_{\nu l}+\omega_p)+i\gamma_{\nu l}]
}\\&
+\frac{\mu_{l\nu}^i\mu_{\nu n}^j\mu_{nm}^k\mu_{ml}^h}{
[(\omega_{nm}+\omega_p+\omega_q+\omega_r)+i\gamma_{nm}]
[(\omega_{\nu m}+\omega_p+\omega_q)+i\gamma_{\nu m}]
[(\omega_{ml}-\omega_p)-i\gamma_{ml}]
}\\&
+\left.\frac{\mu_{l\nu}^h\mu_{\nu n}^i\mu_{nm}^j\mu_{ml}^k}{
[(\omega_{ml}+\omega_p+\omega_q+\omega_r)+i\gamma_{ml}]
[(\omega_{nl}+\omega_p+\omega_q)+i\gamma_{nl}]
[(\omega_{\nu l}+\omega_p)+i\gamma_{\nu l}]
}\right\},
\end{split}
\end{equation}
\end{widetext}
where $\mathcal P_I$ denotes permutations of the frequencies
$\omega_p$, $\omega_q$ and $\omega_r$ with the Cartesian indices
$h,i,k$ permuted simultaneously.

As it was mentioned in the main article, every term in the expressions
for $\chi^{(1)}$ and $\chi^{(3)}$ can be seen as a sum over all
transitions through the intermediate virtual states
\cite{boyd08:book}, with the initial and final state being the same.
Since we do not consider effects of finite temperatures here, the
initial populations are taken in the form:
\begin{equation}
  \label{eq:rho0_metal}
  \rho^{(0)}_{ll}=
  \begin{cases}
    1,&\text{if $\mathcal E\le\mathcal E_F$;}\\
    0,&\text{if $\mathcal E>\mathcal E_F$.}
  \end{cases}
\end{equation}
Moreover, in \refeq{eq:chi3}, due to the Pauli principle, we keep only the transitions over the
intermediate virtual levels which are outside of the ``Fermi sea'', that
is, with $\mathcal E>\mathcal E_F + \mathcal E_0$, where
$\mathcal E_0$ is the ground state. In \refeq{eq:chi1}, in contrast to
\refeq{eq:chi3}, this pre-selection happens automatically. To take
into account the finite depth of our potential, we also do not
consider levels with
$\mathcal E > \mathcal E_W + \mathcal E_F + \mathcal E_0$, where
$\mathcal E_W$ is the work function. For this article, we have taken
$\mathcal E_W=5.1$ eV, $\mathcal E_F=5.53$ eV.

For practical computations, in \refeq{eq:chi1} and \refeq{eq:chi3} we
use the radial parts of the dipole moments given by \refeq{eq:3}, and
average over angular parts \cite{hache86}.
This is possible if assuming that only the transitions with $l\gg 1$ are
relevant, which is indeed the case even for smallest diameters we
consider, as one can see in Fig. 1 of the main article. In this situation,
averaging over the angular dependencies for $-l \le m \le l$ gives
\cite{hache86} the constant factor $\mathcal A = 1/3$ for the linear
susceptibility $\chi^{(1)}$ and $\mathcal A' = 2/15$ for third order
susceptibility $\chi^{(3)}$. Furthermore, when calculating $\chi^{(3)}$, we take
into account
a population-induced correction factor $T_1/T_2$ (see
\cite{hache86}). 
As a result, the susceptibilities obtained in \refeq{eq:chi1},
\refeq{eq:chi3} are corrected as:
\begin{equation}
  \label{eq:chi-corr}
  \chi^{(1)}\to \mathcal A\chi^{(1)}, \,\,
\chi^{(3)}\to \frac{T_1}{T_2}\mathcal A'\chi^{(3)}.
\end{equation}

\subsection{Linear susceptibility for different diameters}

Whereas in Fig. 1(a) of the main article only two particular examples
of the linear susceptibility for spherical nanoparticles are shown, in
\reffig{fig:lin}(a,b) more examples are given, to  illustrate further
the dependence of the super-resonance position on the particle size.

\begin{figure*}[thb]
\includegraphics[width=\textwidth]{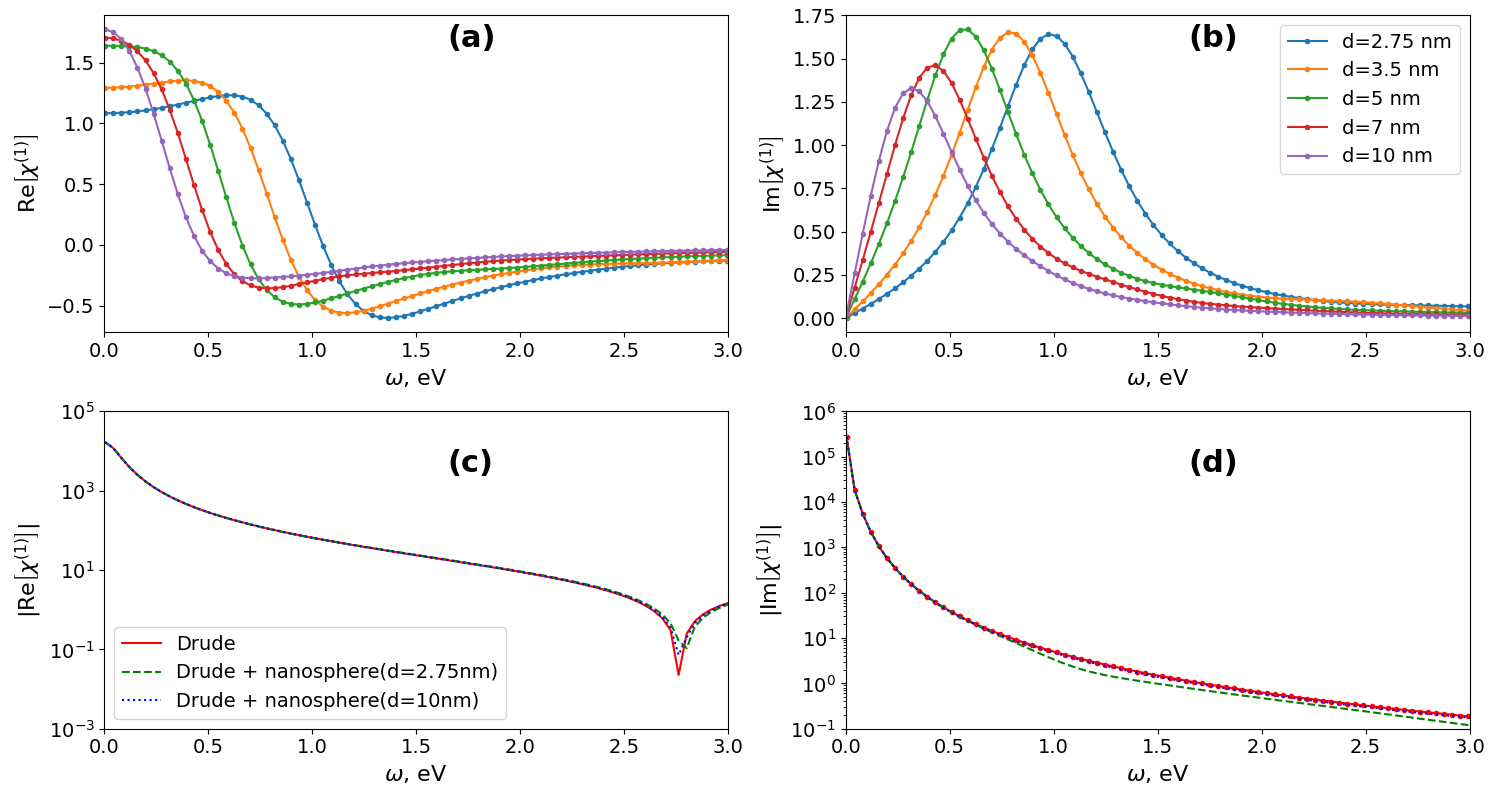}
\caption{ \label{fig:lin} (a,b) Linear susceptibility due to
  confinement-based resonances only, in dependence on frequency for
  spherical nanoparticle of different diameters. (c,d) The Drude
  contribution as well as the full linear susceptibility (Drude +
  confinement) for few  selected diameters. }
\end{figure*}

\subsection{Influence of other mechanisms}
\label{sec:other-mechs}

Although the approaches presented in our article is rather universal
and largely material-independent, in the main paper we, to be
specific, considered the parameters of gold as our basic case since
gold has the most practical importance. However, gold is quite a
``complicated'' metal in the sense that many other mechanisms
contribute to nonlinearity. Besides, in all metals, in addition to
quantum-confinement (billiard-part) the Drude part of $\chi^{(1)}$ is
contributing. In this section we will clarify in more detail the
question how other mechanisms influence the overall nonlinear and
linear response.

\subsubsection{Drude part of $\chi^{(1)}$}
\label{sec:drude-part}

In the Figure 2 of the main article we show the linear response
without the influence of the Drude part of $\chi^{(1)}$ (the first
part in the left side of \refeq{eq:chi1} above, also described by
\refeq{eq:drude} above). The influence of the Drude part is much
larger that the confinement-based part, especially at small
frequencies in MIR and THz range. In particular, \reffig{fig:lin}(c,d)
shows the Drude part alone and together with the quantum confinement
part of the linear susceptibility for few particular diameters. One
can clearly see that the Drude part absolutely dominates in the linear
response at low frequencies. However, this does not mean that that the
confinement-based super-resonance considered in this paper
``disappears'' as we take into account Drude. It can be still
deconvoluted and separated from the Drude part \cite{wyrwas07}.
Besides, at it was shown in the main text, it is directly visible when
considering the nonlinear optical properties such as Kerr effect.

\subsubsection{Effect of interband transitions on the nonlinear response}
\label{sec:effect-bandgap}

In the main article, we considered rather simplified bandgap
consisting only of one band. Whereas such approximation is very well
suitable for some materials such as alkali metals, for other materials
such as gold it could be claimed to be a rather bad approximation. However, at
least in the particular case of gold, as soon as we consider
low-frequency response, it can be shown that the nonlinearity due to
interband transitions is much smaller than the one due to the
billiard resonances.

In the case of gold \cite{christensen71}, there is a strong resonance
in the optical response around 2.4 eV, responsible to the transitions
from the 5d valence band to the 6sp conduction band, as well as a
number of less pronounced resonances in the range between around 2 to
10 eV. That is, the effects of the interband transitions might be
pronounced for the photon energies above 1 eV. Even at the frequency
resonant with the interband transition the impact of confinement-based
resonances to the rectification-like FWM processes we consider is
one or two orders of magnitude larger that the effect of the interband
transitions, as we will see in the next paragraphs.

Let us first consider the Kerr nonlinearity. Experimentally measured
Kerr susceptibility for the photon energy around 2.3 eV, that is,
close to resonance of the above mentioned transition is (for short,
100-fs-scale pulses)
$\chi^{{(3)}}(\omega;\omega,\omega,-\omega)\sim 10^{-18}$ m$^2$/V$^2$
(see for instance \cite{boyd14}; note that in many other references
long, picosecond pulses are considered, demonstrating higher
nonlinearity as discussed in the subsection below). These
measurements, of course, include all effects simultaneously, in
particular intraband transitions and confinement-induced effects. We
see that our calculations give the same order of magnitude of
susceptibility for this frequency [see Fig. 3(a) of the main article]
with only the confinement-based nonlinearity included. This means that
the interband transitions do not dominate the intraband even at the
interband resonance frequency; the impact of confinement-based
resonances is of the same order of magnitude or higher.

On the other hand, as we decrease the frequency from the interband
resonance towards THz range, the influence of the interband resonance
quickly decreases whereas the influence of the confinement-based
resonance increases [see Fig. 3(a) of the main article]. Therefore, we
come to the conclusion that in the THz range the Kerr nonlinearity
$\chi^{{(3)}}(\omega;\omega,\omega,-\omega)$ is indeed dominated by
the confinement-based (intraband) transitions.

The same is also true for the FWM processes responsible for
rectification-like effects considered in the main article: the
influence of the interband transitions must be significantly smaller than
of the intraband (confinement-based) ones. In order to make our
estimation more quantitative at this point, we use the estimation
technique described in \cite{hache88}. We start from the Kerr process
$\chi^{{(3)}}(\omega;\omega,\omega,-\omega)$ assuming it to be fully
resonant to the interband transition (``worst-case scenario,'' where
the interband action is maximal). We
consider then the processes
$\chi^{{(3)}}(\delta;\omega+\delta,\omega,-2\omega)$ and 
$\chi^{{(3)}}(\delta;\omega,\omega,-2\omega+\delta)$. Because of the
missing resonant terms in \refeq{eq:chi3} (which gives a factor $\sim
\omega^2$ in comparison to the interband resonant case) and
also because of smaller population of the virtual levels (factor
$\sim T^2_2$), the  nonlinear susceptibility is reduced by a factor
$(\omega T_2)^2\sim 10^2$ comparing to the Kerr susceptibility at the interband resonance. Since, as it was established before, even
at the interband resonance the confinement-based Kerr nonlinearity is
at least of the same order of magnitude that the interband-induced
Kerr nonlinearity, we therefore conclude that for the FWM processes the
intraband (confinement-based) resonances are at least by the factor of
100 larger than the interband ones. Based on our calculations of the
intraband nonlinearities [see Fig. 3(b) of the main article], we can
estimate the interband effect to the nonlinear susceptibility for the considered FWM processes as
10$^{-20}$-10$^{-21}$ m$^2$/V$^2$ for $\omega$ at the interband
resonance (and even lower away from that resonance). This is much less
than the confinement-based impact as shown in Fig. 3 of the main article.

\begin{figure*}[thb]
\includegraphics[width=\textwidth]{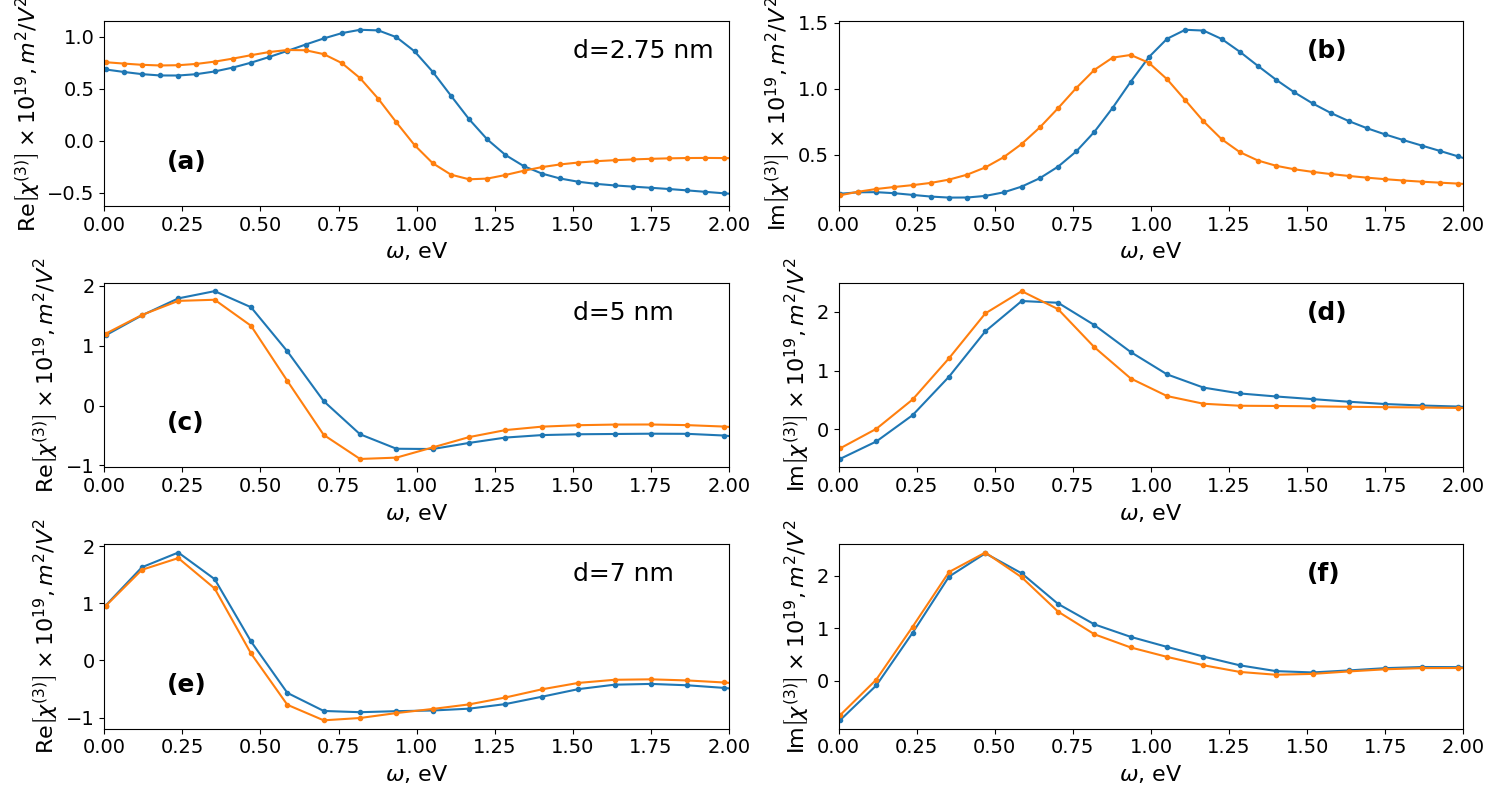}
\caption{ \label{fig:brillouin} Real (a,c,e) and imaginary (b,d,f)
  part of the nonlinear susceptibility
  $\chi^{(3)}(\delta;\omega,\omega,-2\omega+\delta)$ for different
  diameters (see legend), corresponding to Fig. 3(b) of the main
  article, effectively taking into account (orange curves) and without
  taking account (blue lines) the interband transitions as introduced
  by \refeq{eq:Einterband}. The latter correspond to Fig. 3(b) of the
  main article. }
\end{figure*}

As an alternative and fully independent method to estimate the impact
of the interband transitions we introduce a gap directly into our
numerical model. That is, we modify our single-band structure as the
following:
\begin{equation}
  \label{eq:Einterband}
 \mathcal{E}_{nl} =  \begin{cases}
      \mathcal{E}_0\alpha_{nl}^2,&\text{if $|\alpha_{nl}|\le\frac{2\pi}{\Lambda}$,}\\
      \mathcal{E}_0\alpha_{nl}^2 + E_g,&\text{if $|\alpha_{nl}|>\frac{2\pi}{\Lambda}$,}
   \end{cases}
\end{equation}
where $\Lambda$ is the lattice constant (for gold $\Lambda \approx 4$
\AA), $E_g\approx 2.4$ eV. This modification mimics the bandgap which
opens near the edges of the Brillouin zone. The comparison of two
calculations, that is, using the single band \refeq{eq:E} model and
two band model \refeq{eq:Einterband}, are shown in
\reffig{fig:brillouin} for different exemplary diameters. One can see
that, whereas the interband transition does provide some limited
modification to the confinement-based dynamics for very small nanoparticles
$d\lessapprox 3$ nm, this influence quickly decreases and becomes
negligible for larger diameters.

\subsubsection{Thermal, hot-electron and related effects}
\label{sec:thermal-effects}

Taking into account temperature introduces several effects, which were
neglected in the main article. First of all, it leads to an additional
hot-electron
contribution into nonlinearly \cite{boyd14}, which can overcome, by
several orders of magnitude, the nonlinearities considered in this
article up to now.
The hot electron mechanism involves laser-induced intraband excitation
in the conduction band, followed by the energy dissipation of the
excited electrons. This process leads to a modification of Fermi-Dirac
distribution which depends on the frequency and intensity of the pump,
leading thereby to frequency-dependent nonlinearity
\cite{voisin00,besteiro19,hartland17}. The key role in the quick
thermalization is played by the electron-electron and electron-phonon
interactions.

This nonlinearity has relatively slow, sub-picosecond-scale, turn-on
time \cite{sun94}, and therefore its influence quickly decreases with
decreasing of the pulse duration \cite{boyd14}. For the pulses
considered here (10-30 femtoseconds) this nonlinearity plays a
negligible role. Indeed, the experimentally measured Kerr nonlinearity
for gold nanospheres \cite{boyd14, rotenberg07} for the pulses of 100
fs duration corresponds, by the order of magnitude ($\sim 10^{-18}$
m$^2$/V$^2$, see also discussion in the previous subsection) to our
calculations at around the same frequency (cf. Fig. 3(a) of the main
article). Since in our calculations we do not take into account the
hot electron nonlinearities, we come to the conclusion that for short
pulses such nonlinearities are pretty much negligible in comparison
to the confinement-based resonances, or at least do not play a
dominant role. This is even more true if we consider lower frequencies
towards THz range, since the confinement-based nonlinearity has a
resonance at low frequencies, whereas the thermal nonlinearity is not
expected to demonstrate a resonant behaviour.

A part of the above-described thermalization process is the
electron-electron interaction. Fast thermalization, indeed, is the
primarily consequence of the electron-electron interactions
\cite{hertel96,hartland17}. Electron-electron interactions are
trackable in the linear properties of the nanostructure (see for
instance \cite{voisin00}), however, the corresponding modification is
rather minor and even this small modification starts to be visible at
the time scale of few tens of femtoseconds.

\begin{figure*}[thb]
\includegraphics[width=\textwidth]{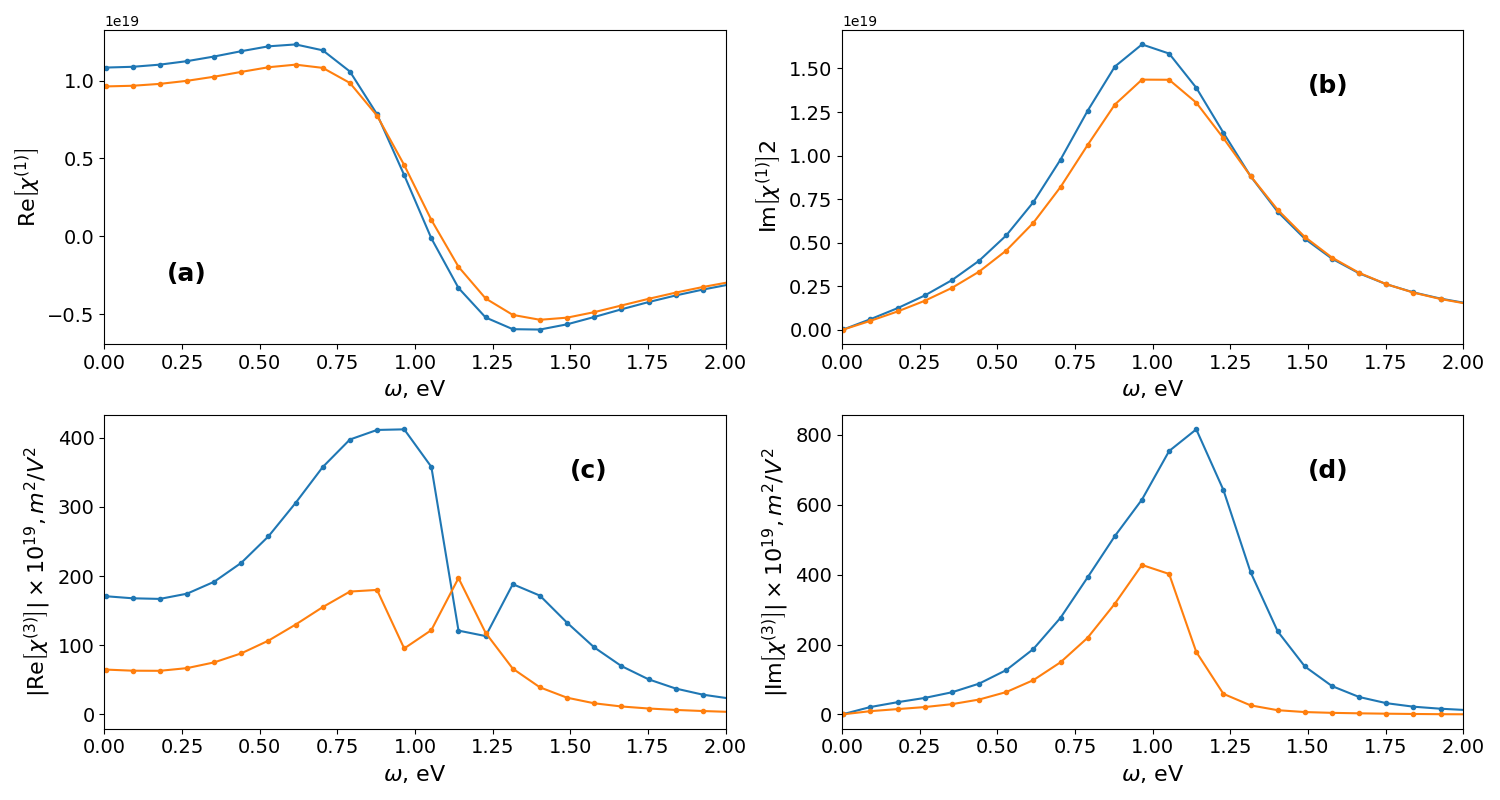}
\caption{ \label{fig:temperature} Real (a,c) and imaginary (b,d)
  part of the linear (a,b) and nonlinear (Kerr) (b,c) susceptibility
   for the structure $d=2.75$ assuming sharp transition at $E=\mathcal
   E_F$ (blue lines) and taking into account Fermi-Dirac distribution
   for $T=300$ K (orange lines). }
\end{figure*}

Another thermal effect is the overall non-rectangular shape of the electron
distribution near the Fermi zone edge as soon as the temperature is
nonzero (in our calculations in the main article we assumed zero
temperature). The effect of nonzero temperature in the vicinity of
Fermi-level is shown in \reffig{fig:temperature} for an exemplary
diameters $d=2.75$ nm and temperature $T=300$ K. It is obtained by
modifying $\rho_{ll}^{(0)}$ from \refeq{eq:rho0_metal} to the
Fermi-Dirac distribution for the finite temperature. One can see that
this modifies only slightly the linear response. The nonlinear
response is also modified quite moderately.

\subsection{Level statistics details}
\label{sec:level-stat-deta}

\subsubsection{General definitions}
\label{sec:general-definitions}

In this section we describe different variants of the level
statistics, extending the discussion related to Fig. 2 of the main
article. To understand the mechanisms, governing the formation of one
single super-resonance it is very constructive to consider the level
statistics, varying the selection rules included into that statistics.
Commonly in the quantum billiard and quantum chaos theory
\cite{stoeckmann99:book} one considers the so called neighboring level
statistics. That is, we consider the difference between the
neighboring levels $\delta \omega_i = \omega_{i+1}-\omega_i$, where
the eigenfrequencies $\omega_i$ are obtained by ordering of the
eigenfrequencies in the increasing order, that is, we order them in
such a way that $\omega_j\le\omega_i$ for $j<i$. In the case of
nanospheres the corresponding eigenvalues are
$\hbar \omega_{nl}=\mathcal E_{nl}$, cf. \refeq{eq:E}, reaaranged
accordingly. Note that $i$, $j$ here are single indices (and not
multiindices as in Sec.~\ref{sec:lin-nonlin-prop}). Then, the
probability  $P(\omega)d \omega$ that $\delta \omega_i$
is located in the range between $\omega$ and $\omega+d\omega$ is calculated.
The easiest way to visualize such statistics is to use the histogram
technique. For nanospheres this traditional neighboring level
statistics is presented in Fig. 2a of the main article (yellow bars).

The neighboring level statistics do allow to determine the universality
classes of different billiards. It, however, does not take into
account the properties of eigenfunctions and of the transition rules,
so it seems to be of little use for the optical properties. In order
to improve usability for optics, we extend the statistics to take into
account all transitions, not only neighboring, and impose
additional selection rules. That is, we consider the statistics of
energy differences $\omega_{ij} = \omega_{i}-\omega_j$ for $i>j$
(assuming the ordering of $\omega_j$ as discussed above). In addition,
when constructing the probability density $P(\omega)$ of $\omega_{ij}$
being in the interval $[\omega,\omega+d\omega]$, we take into account
the ``strength'' of the transition by weighting $P(\omega_{ij})$ with
a weight
\begin{equation}
w_{ij}=\Delta_{ij}|\mu_{ij}|^2, \label{eq:w_ij}
\end{equation}
where $\mu_{ij}$ is the dipole momentum of the corresponding transition
$i\to j$, and $\Delta_{ij}$ is defined as
\begin{equation}
  \label{eq:Delta_ij}
  \Delta_{ij}=
  \begin{cases}
    1,&\text{if $\hbar \omega_i \le \mathcal E_F$ and $\hbar \omega_j > \mathcal E_F$}\\
    0,&\text{otherwise,}
  \end{cases}
\end{equation}
that is, takes into account the Fermi-Dirac distribution
\refeq{eq:rho0_metal} (we call it below ``the Fermi sea condition''). The
resulting statistics is shown in Fig. 2 (red and blue bars) and
repeated for convenience as the inset in \reffig{fig:stat}(a) -- for
the larger  frequency range. We note that the traditional statistics
discussed in the previous paragraph is a partial case of this more
general approach; namely, we obtain the traditional neighboring level
statistics assuming $w_{ij}=\delta_{i,j=i+1}$.

\subsubsection{Natural shape and width of the super-resonance for
  nanospheres}
\label{sec:natural-shape-width}

To find out the ``natural width'' of the super-resonance for the case
of nanostructures, we use a more more precise version of \refeq{eq:alpha_nl_approx}:
\begin{equation}
    \alpha_{nl}=(2n+l+s_{nl})\pi/2,
\end{equation}
which includes corrections $s_{nl}$ to the values of the roots. All
transitions contributing to the super-resonance are characterized by
the same value of $2n+l$ before and after the transition, that is,
$2n'+l'=2n+l+1$. However, values of $n$ and $l$ can be different.
Therefore the energy difference between the eigenfunctions
$\ket{\psi_{nl}}$ and $\ket{\psi_{n'l'}}$ will be modified as follows:
\begin{widetext}
\begin{equation}
    \Delta \mathcal E=\frac{\pi^2\mathcal E_0}{4}\left[(2n'+l'+s_{n'l'})^2-(2n+l+s_{nl})^2\right]\simeq\frac{\pi^2\mathcal E_0}{2}(2n+l)(1+s_{n'l'}-s_{nl}).
\end{equation}
\end{widetext}
For the typical range of $n$ and $l$ actual for our particular
situation, we approximate $s_{nl}$ as being distributed in the range
[-0.25,0]. For the probability distribution of the difference
$s_{nl}-s_{n'l'}$, given by
$P(s_{n'l'}-s_{nl})\sim\int P(s)P(s+s_{n'l'}-s_{nl})ds$, we obtain a
triangular symmetric shape with the maximum at zero and full width of
$1/2$, that is we have a constant FWHM of $1/4$ of the distribution of
the difference $s_{nl}-s_{n'l'}$. Using \refeq{eq:DE} we obtain that
this translates to the width $\approx \omega_{\mathrm{conf}}/4$ in
frequency, where $\omega_{\mathrm{conf}}$ is the position of the
super-resonance as defined in Eq. 1 of the main article. If we
consider the dipole-momentum-weighted statistics as described below,
this simplified conclusion will be modified by the fact that the
dipole momentum depends on the energy level difference. Such
dependence will result in sharper lower-frequency shoulder of the
statistics and smoother higher-frequency shoulder. Both such shape and
the width of the peak in the statistics are in a surprisingly good
agreement with the findings shown in Fig. 2 of the main article (for
the case of large T$_2$). Of course, reducing the $T_2$ to
room-temperature values additionally increases the width of the
super-resonance beyond the natural width of
$\approx \omega_{\mathrm{conf}}/4$.


\subsubsection{Influence of different mechanisms on the statistics}
\label{sec:infl-diff-mech}

\begin{figure*}[thb]
\includegraphics[width=\textwidth]{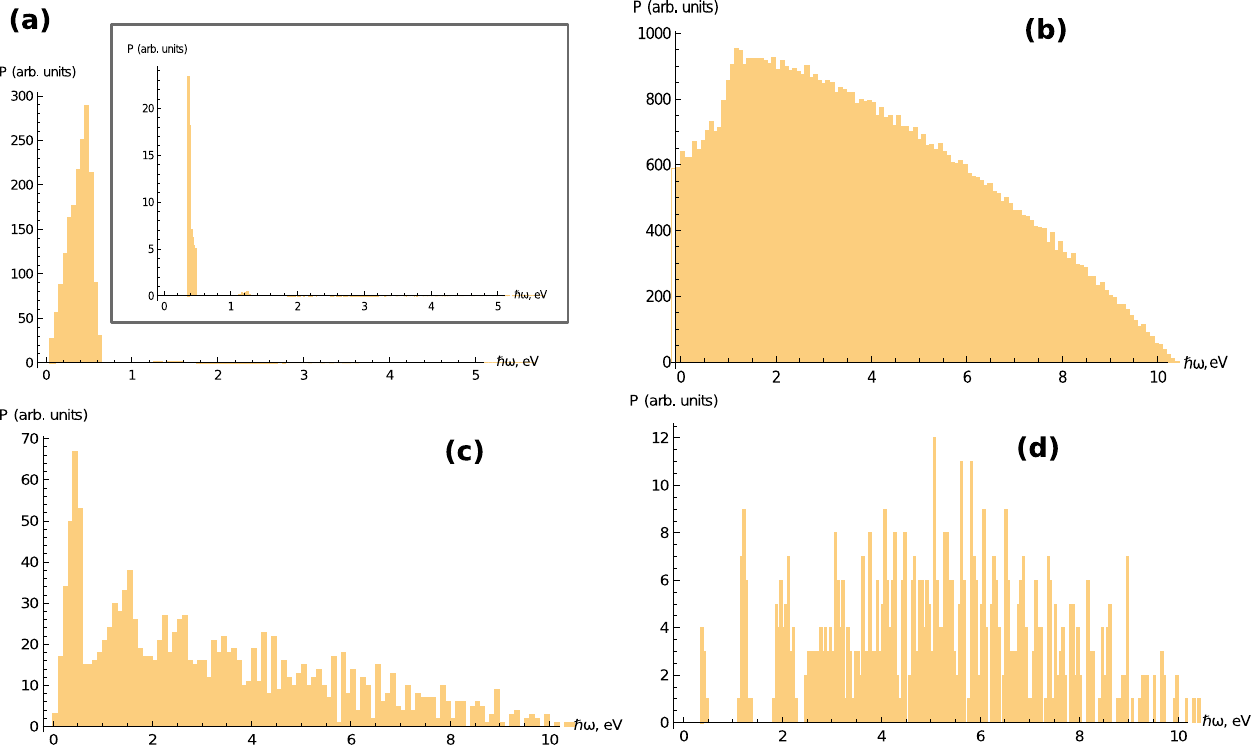}
\caption{ \label{fig:stat} Different variants of level statistics. (a)
  $P(\omega_{ij})$ weighted as given by \refeq{eq:w_ij}, only with
  $\Delta_{ij}=1$, that is not taking into account Fermi sea condition
  (delivering thereby the statistics for dielectrics). Inset to (a):
  Repetition of Fig. 1 of the main article on the larger energy scale,
  that is: $P(\omega_{ij})$ weighted as given by \refeq{eq:w_ij}. (b)
  The same as in (a) but assuming in addition $\mu_{ji}=1$, that is,
  unweighted. (c) The same as in (b) but assuming $w_{ji}=0$ if
  $\mu_{ij}=0$ and $w_{ij}=1$ otherwise. (d) The same as in (c) but
  taking into account the Fermi sea restriction \refeq{eq:Delta_ij}. }
\end{figure*}

Varying the selection rules incorporated into the weighting
\refeq{eq:w_ij}, we can study how these rules influence the
statistics. Different variants of statistics are shown in
\reffig{fig:stat}. In particular, in \reffig{fig:stat}(a) we consider
$P(\omega_{ij})$ weighted by \refeq{eq:w_ij} with $\Delta_{ij}=1$,
that is not taking into account the Fermi sea condition -- i.e. all
transition, not only the transitions from the below to above the Fermi
sea level, are allowed. This situation describes dielectrics rather
than metals, and should be contrasted to the ``metallic case'', that
is, the situation where also the ``Fermi sea condition'' is satisfied
[inset to \reffig{fig:stat}(a) as well as Fig. 2a of the main
article]. One can see that in the former case the resonance is much
more broad and, in addition, noticeably shifted to the lower
frequencies.

On the other hand, if we remove all restrictions at all, that is,
consider $w_{ij}=1$ for all $i$, $j$, we will see very broad
distribution over the scale of many eV \reffig{fig:stat}(b). The
statistics changes not too significantly if we take into account only
allowed transitions (but not yet distinguishing between the strengths
of the transitions, that is, assuming $w_{ji}=0$ if $\mu_{ij}=0$ and
$w_{ij}=1$ otherwise, and also not taking into account the Fermi sea
condition). Yet, in this case the peak, corresponding to the
low-frequency super-resonance (at around 0.5 eV) does already appear
[see \reffig{fig:stat}(c)].
This peak becomes even sharper if we in addition take into account the
Fermi sea condition as shown in \reffig{fig:stat}(d). But, in addition
to this low-frequency peak, in \reffig{fig:stat}(d) we see also many
other peaks at higher frequency. These many peaks disappear as we take
into account the ``strengths'' of the transitions ($\mu^2$-weighting as
given by \refeq{eq:w_ij}), see inset to \reffig{fig:stat}(a) and Fig.
2a of the main article.

Therefore, we conclude that there is, in general, a lot of possible
confinement-based transitions at low and high frequencies. The key
role in the formation of a single super-resonance is played by the
dipole moments, that is, by the symmetry and composition of the
wavefunctions, and only to a lesser extent by the Fermi sea condition.
As we take them into account, we are left with a single bunch of
closely spaced resonances, which, taking into account that every of
these resonances are broadband, merge into a super-resonance. Whereas
the consideration here was focused on the case of nanospheres, we
remark that, most probably, this particular situation is quite
geometry-independent, at least for regular billiards. Indeed, in this
case we expect that distantly spaced eigenfunctions have very
different number of oscillations in every spatial direction, making
$\mu_{ij}$ small. This, however, is not necessarily the case for the
irregular, chaotic billiards which have also rather chaotic
eigenfunctions. Finally, as Fig. 2(b) of the main article shows, we
can tune the positions of the resonances in such a way that the
resulting super-resonance is broadened.

\subsection{Effective properties of nanocomposite}

The effective linear properties of the nanocomposite for arbitrary
frequency $\omega$ are calculated  using the effective medium approach
\cite{zeng88-effective-n-n2-nanoparticles} using the linear properties of the
host $\varepsilon_h(\omega)$ (in our case SiO$_2$), and nanostructures
$\varepsilon_\mathrm{NP}(\omega)$ [which is given by
$\varepsilon_\mathrm{NP}(\omega) = 1+ \chi^{(1)}$, where $\chi^{(1)}$
is calculated according to \refeq{eq:chi1}], as:
\begin{equation}
  \label{eq:lin_eps_eff}
  \varepsilon_{\mathrm{eff}}(\omega) = \varepsilon_h(\omega) + f x(\omega)
  (\varepsilon_\mathrm{NP}(\omega) - \varepsilon_h(\omega)).
\end{equation}
where $f$ is the filling factor, $x(\omega)$ is defined as
\begin{equation}
  \label{eq:x}
  x(\omega) = \frac{3\varepsilon_{h}(\omega)}{\varepsilon_{\mathrm{NP}}(\omega)+ 2\varepsilon_h(\omega)}.
\end{equation}
We note that at Mie resonance $|x|$ is especially large. Finally, the
effective nonlinear susceptibility $\chi^{(3)}_{\mathrm{eff}}$ for
every process is calculated for given linear and nonlinear properties
of the nanostructures and host as follows:
\begin{equation}
  \label{eq:chi3eff}
  \chi^{(3)}_{\mathrm{eff}} =   f \chi^{(3)}_\mathrm{NP}x(\omega_0)x_(\omega_1)^2x(\omega_2),
\end{equation}
where $\chi^{(3)}_\mathrm{NP}$ is the nonlinear susceptibility given by
Eq. 5  or Eq. 6 of the main article, with $\chi^{(3)}$ in
those expressions calculated using \refeq{eq:chi3}. 

\subsection{Propagation equations}
\label{sec:prop-equat}

Assuming slowly varying envelope approximation and neglecting
nonlinear effects for the pump waves, the governing equations are:
\begin{gather}
  \label{eq:prop_dz}
  \frac{\partial A_0}{\partial z} =
  -\frac{1}{2c\varepsilon_0}\frac{\partial P}{\partial t} =
  -\frac{i\delta \chi^{(3)}_{\mathrm{eff}}(\delta)}{2c} A_1^2(z)A^*_2(z) -
  \alpha_0 A_0,\\
  \partial_z A_n = ik_n A_n - \alpha_n A_n,\, n=1,2,
  \label{eq:prop_dz_1}
\end{gather}
where $c$ is the speed of light in vacuum, $\varepsilon_0$ is vacuum
permittivity, $A_i$, $\alpha_i$ $k_i$, $i=0,1,2$ are correspondingly
the slow (complex) amplitudes, linear losses, and wavevectors for
$n$th harmonic (here signal is assumed to be the ``0th harmonic''),
and $\chi^{(3)}_{\mathrm{eff}}(\delta)$ is the effective nonlinear
susceptibility for the corresponding process.
\refeqs{eq:prop_dz}{eq:prop_dz_1} allows an analytical solution, given by:
\begin{equation}
  \label{eq:prop_sol}
  A_0 = -i A e^{-\alpha_0L} (e^{\kappa L} - 1)/\kappa,
\end{equation}
where $A=\delta \chi^{(3)}_\mathrm{eff} A_1^2(0)A_2^*(0)/2c$, $L$ is the
propagation distance, $\kappa = i(k_0 + 2k_1 - k_2) - 2\alpha_1 - \alpha_2 + \alpha_0$. 

Assuming the pulse durations of around 30 fs, we must consider two
regions for the pump where conversion works significantly differently.
For the signal in the THz range ($\delta/2\pi \le$ 30 THz), the
frequencies $j\omega_0$ and $j\omega_0+\delta$ ($j=1,2$) are both
located within the spectrum of the pump. The nonlinearity in this case
is driven by three contribution types mentioned above, and must be
considered as the following sum:
\begin{multline}
\chi^{(3)}_{\mathrm{NP}}
(\delta) \approx
\chi^{(3)}(\delta;\omega_0,\omega_0,-2\omega_0+\delta) \\
+ \chi^{(3)}(\delta;\omega_0+\delta,\omega_0,-2\omega_0) +\chi^{(3)}(\delta;\omega_0,\omega_0+\delta,-2\omega_0).\label{eq:chi3_np1}
\end{multline}

In contrast, for the signal in MIR range $\delta/2\pi >$ 30 THz, the
components $\omega_0+\delta$, $2\omega_0+\delta$ are not within the
pump spectrum anymore. In this case, in order to make the conversion
efficient, we must, for instance, shift the second harmonic:
$2\omega_0\to2\omega_0+\delta$. The only effective nonlinear process
in this case is 
\begin{equation}
\chi^{(3)}_{\mathrm{NP}}(\delta) =
\chi^{(3)}(\delta;\omega_0,\omega_0,-2\omega_0+\delta).\label{eq:chi3_np2}
\end{equation}



\begin{thebibliography}{55}%
\makeatletter
\providecommand \@ifxundefined [1]{%
 \@ifx{#1\undefined}
}%
\providecommand \@ifnum [1]{%
 \ifnum #1\expandafter \@firstoftwo
 \else \expandafter \@secondoftwo
 \fi
}%
\providecommand \@ifx [1]{%
 \ifx #1\expandafter \@firstoftwo
 \else \expandafter \@secondoftwo
 \fi
}%
\providecommand \natexlab [1]{#1}%
\providecommand \enquote  [1]{``#1''}%
\providecommand \bibnamefont  [1]{#1}%
\providecommand \bibfnamefont [1]{#1}%
\providecommand \citenamefont [1]{#1}%
\providecommand \href@noop [0]{\@secondoftwo}%
\providecommand \href [0]{\begingroup \@sanitize@url \@href}%
\providecommand \@href[1]{\@@startlink{#1}\@@href}%
\providecommand \@@href[1]{\endgroup#1\@@endlink}%
\providecommand \@sanitize@url [0]{\catcode `\\12\catcode `\$12\catcode
  `\&12\catcode `\#12\catcode `\^12\catcode `\_12\catcode `\%12\relax}%
\providecommand \@@startlink[1]{}%
\providecommand \@@endlink[0]{}%
\providecommand \url  [0]{\begingroup\@sanitize@url \@url }%
\providecommand \@url [1]{\endgroup\@href {#1}{\urlprefix }}%
\providecommand \urlprefix  [0]{URL }%
\providecommand \Eprint [0]{\href }%
\providecommand \doibase [0]{http://dx.doi.org/}%
\providecommand \selectlanguage [0]{\@gobble}%
\providecommand \bibinfo  [0]{\@secondoftwo}%
\providecommand \bibfield  [0]{\@secondoftwo}%
\providecommand \translation [1]{[#1]}%
\providecommand \BibitemOpen [0]{}%
\providecommand \bibitemStop [0]{}%
\providecommand \bibitemNoStop [0]{.\EOS\space}%
\providecommand \EOS [0]{\spacefactor3000\relax}%
\providecommand \BibitemShut  [1]{\csname bibitem#1\endcsname}%
\let\auto@bib@innerbib\@empty
\bibitem [{\citenamefont {Mie}(1908)}]{mie1908}%
  \BibitemOpen
  \bibfield  {author} {\bibinfo {author} {\bibfnamefont {Gustav}\ \bibnamefont
  {Mie}},\ }\bibfield  {title} {\enquote {\bibinfo {title} {{Beitr{\"a}ge zur
  Optik tr{\"u}ber Medien, speziell kolloidaler Metall{\"o}sungen}},}\
  }\href@noop {} {\bibfield  {journal} {\bibinfo  {journal} {Annalen der
  physik}\ }\textbf {\bibinfo {volume} {330}},\ \bibinfo {pages} {377--445}
  (\bibinfo {year} {1908})}\BibitemShut {NoStop}%
\bibitem [{\citenamefont {Kreibig}\ and\ \citenamefont
  {Vollmer}(2013)}]{kreibig13:book}%
  \BibitemOpen
  \bibfield  {author} {\bibinfo {author} {\bibfnamefont {Uwe}\ \bibnamefont
  {Kreibig}}\ and\ \bibinfo {author} {\bibfnamefont {Michael}\ \bibnamefont
  {Vollmer}},\ }\href@noop {} {\emph {\bibinfo {title} {Optical properties of
  metal clusters}}},\ Vol.~\bibinfo {volume} {25}\ (\bibinfo  {publisher}
  {Springer Science \& Business Media},\ \bibinfo {year} {2013})\BibitemShut
  {NoStop}%
\bibitem [{\citenamefont {Kim}\ \emph {et~al.}(2010)\citenamefont {Kim},
  \citenamefont {Husakou},\ and\ \citenamefont {Herrmann}}]{kim10c}%
  \BibitemOpen
  \bibfield  {author} {\bibinfo {author} {\bibfnamefont {Kwang-Hyon}\
  \bibnamefont {Kim}}, \bibinfo {author} {\bibfnamefont {Anton}\ \bibnamefont
  {Husakou}}, \ and\ \bibinfo {author} {\bibfnamefont {Joachim}\ \bibnamefont
  {Herrmann}},\ }\bibfield  {title} {\enquote {\bibinfo {title} {Linear and
  nonlinear optical characteristics of composites containing metal
  nanoparticles with different sizes and shapes},}\ }\href
  {http://www.opticsexpress.org/abstract.cfm?URI=oe-18-7-7488} {\bibfield
  {journal} {\bibinfo  {journal} {Opt. Express}\ }\textbf {\bibinfo {volume}
  {18}},\ \bibinfo {pages} {7488--7496} (\bibinfo {year} {2010})}\BibitemShut
  {NoStop}%
\bibitem [{\citenamefont {Kawabata}\ and\ \citenamefont
  {Kubo}(1966)}]{kawabata66}%
  \BibitemOpen
  \bibfield  {author} {\bibinfo {author} {\bibfnamefont {Arisato}\ \bibnamefont
  {Kawabata}}\ and\ \bibinfo {author} {\bibfnamefont {Ryogo}\ \bibnamefont
  {Kubo}},\ }\bibfield  {title} {\enquote {\bibinfo {title} {Electronic
  properties of fine metallic particles. {II}. {P}lasma resonance
  absorption},}\ }\href@noop {} {\bibfield  {journal} {\bibinfo  {journal} {J.
  Phys. Soc. Japan}\ }\textbf {\bibinfo {volume} {21}},\ \bibinfo {pages}
  {1765--1772} (\bibinfo {year} {1966})}\BibitemShut {NoStop}%
\bibitem [{\citenamefont {Ruppin}\ and\ \citenamefont
  {Yatom}(1976)}]{ruppin76}%
  \BibitemOpen
  \bibfield  {author} {\bibinfo {author} {\bibfnamefont {R.}~\bibnamefont
  {Ruppin}}\ and\ \bibinfo {author} {\bibfnamefont {H.}~\bibnamefont {Yatom}},\
  }\bibfield  {title} {\enquote {\bibinfo {title} {Size and shape effects on
  the broadening of the plasma resonance absorption in metals},}\ }\href@noop
  {} {\bibfield  {journal} {\bibinfo  {journal} {Phys. Status Solidi B}\
  }\textbf {\bibinfo {volume} {74}},\ \bibinfo {pages} {647--654} (\bibinfo
  {year} {1976})}\BibitemShut {NoStop}%
\bibitem [{\citenamefont {Kraus}\ and\ \citenamefont {Schatz}(1983)}]{kraus83}%
  \BibitemOpen
  \bibfield  {author} {\bibinfo {author} {\bibfnamefont {W.~A.}\ \bibnamefont
  {Kraus}}\ and\ \bibinfo {author} {\bibfnamefont {George~C.}\ \bibnamefont
  {Schatz}},\ }\bibfield  {title} {\enquote {\bibinfo {title} {Plasmon
  resonance broadening in small metal particles},}\ }\href@noop {} {\bibfield
  {journal} {\bibinfo  {journal} {J. Chem. Phys.}\ }\textbf {\bibinfo {volume}
  {79}},\ \bibinfo {pages} {6130--6139} (\bibinfo {year} {1983})}\BibitemShut
  {NoStop}%
\bibitem [{\citenamefont {Amendola}\ \emph {et~al.}(2017)\citenamefont
  {Amendola}, \citenamefont {Pilot}, \citenamefont {Frasconi}, \citenamefont
  {Marag{\`o}},\ and\ \citenamefont {Iat{\`\i}}}]{amendola17}%
  \BibitemOpen
  \bibfield  {author} {\bibinfo {author} {\bibfnamefont {Vincenzo}\
  \bibnamefont {Amendola}}, \bibinfo {author} {\bibfnamefont {Roberto}\
  \bibnamefont {Pilot}}, \bibinfo {author} {\bibfnamefont {Marco}\ \bibnamefont
  {Frasconi}}, \bibinfo {author} {\bibfnamefont {Onofrio~M}\ \bibnamefont
  {Marag{\`o}}}, \ and\ \bibinfo {author} {\bibfnamefont {Maria~Antonia}\
  \bibnamefont {Iat{\`\i}}},\ }\bibfield  {title} {\enquote {\bibinfo {title}
  {Surface plasmon resonance in gold nanoparticles: a review},}\ }\href@noop {}
  {\bibfield  {journal} {\bibinfo  {journal} {J. Phys. Condens. Matter}\
  }\textbf {\bibinfo {volume} {29}},\ \bibinfo {pages} {203002} (\bibinfo
  {year} {2017})}\BibitemShut {NoStop}%
\bibitem [{\citenamefont {Scholl}\ \emph {et~al.}(2012)\citenamefont {Scholl},
  \citenamefont {Koh},\ and\ \citenamefont {Dionne}}]{scholl}%
  \BibitemOpen
  \bibfield  {author} {\bibinfo {author} {\bibfnamefont {J.}~\bibnamefont
  {Scholl}}, \bibinfo {author} {\bibfnamefont {A.}~\bibnamefont {Koh}}, \ and\
  \bibinfo {author} {\bibfnamefont {J.}~\bibnamefont {Dionne}},\ }\bibfield
  {title} {\enquote {\bibinfo {title} {Quantum plasmon resonances of individual
  metallic nanoparticles},}\ }\href@noop {} {\bibfield  {journal} {\bibinfo
  {journal} {Nature}\ }\textbf {\bibinfo {volume} {483}},\ \bibinfo {pages}
  {421–427} (\bibinfo {year} {2012})}\BibitemShut {NoStop}%
\bibitem [{\citenamefont {Morton}\ \emph {et~al.}(2011)\citenamefont {Morton},
  \citenamefont {Silverstein},\ and\ \citenamefont {Jensen}}]{morton11}%
  \BibitemOpen
  \bibfield  {author} {\bibinfo {author} {\bibfnamefont {Seth~M}\ \bibnamefont
  {Morton}}, \bibinfo {author} {\bibfnamefont {Daniel~W}\ \bibnamefont
  {Silverstein}}, \ and\ \bibinfo {author} {\bibfnamefont {Lasse}\ \bibnamefont
  {Jensen}},\ }\bibfield  {title} {\enquote {\bibinfo {title} {Theoretical
  studies of plasmonics using electronic structure methods},}\ }\href@noop {}
  {\bibfield  {journal} {\bibinfo  {journal} {Chem. Rev.}\ }\textbf {\bibinfo
  {volume} {111}},\ \bibinfo {pages} {3962--3994} (\bibinfo {year}
  {2011})}\BibitemShut {NoStop}%
\bibitem [{\citenamefont {Philip}\ \emph {et~al.}(2012)\citenamefont {Philip},
  \citenamefont {Chantharasupawong}, \citenamefont {Qian}, \citenamefont
  {Jin},\ and\ \citenamefont {Thomas}}]{philip12}%
  \BibitemOpen
  \bibfield  {author} {\bibinfo {author} {\bibfnamefont {Reji}\ \bibnamefont
  {Philip}}, \bibinfo {author} {\bibfnamefont {Panit}\ \bibnamefont
  {Chantharasupawong}}, \bibinfo {author} {\bibfnamefont {Huifeng}\
  \bibnamefont {Qian}}, \bibinfo {author} {\bibfnamefont {Rongchao}\
  \bibnamefont {Jin}}, \ and\ \bibinfo {author} {\bibfnamefont {Jayan}\
  \bibnamefont {Thomas}},\ }\bibfield  {title} {\enquote {\bibinfo {title}
  {Evolution of nonlinear optical properties: from gold atomic clusters to
  plasmonic nanocrystals},}\ }\href@noop {} {\bibfield  {journal} {\bibinfo
  {journal} {Nano Lett.}\ }\textbf {\bibinfo {volume} {12}},\ \bibinfo {pages}
  {4661--4667} (\bibinfo {year} {2012})}\BibitemShut {NoStop}%
\bibitem [{\citenamefont {Boyd}\ \emph {et~al.}(2014)\citenamefont {Boyd},
  \citenamefont {Shi},\ and\ \citenamefont {De~Leon}}]{boyd14}%
  \BibitemOpen
  \bibfield  {author} {\bibinfo {author} {\bibfnamefont {Robert~W}\
  \bibnamefont {Boyd}}, \bibinfo {author} {\bibfnamefont {Zhimin}\ \bibnamefont
  {Shi}}, \ and\ \bibinfo {author} {\bibfnamefont {Israel}\ \bibnamefont
  {De~Leon}},\ }\bibfield  {title} {\enquote {\bibinfo {title} {The third-order
  nonlinear optical susceptibility of gold},}\ }\href@noop {} {\bibfield
  {journal} {\bibinfo  {journal} {Opt. Commun.}\ }\textbf {\bibinfo {volume}
  {326}},\ \bibinfo {pages} {74--79} (\bibinfo {year} {2014})}\BibitemShut
  {NoStop}%
\bibitem [{\citenamefont {Qian}\ \emph {et~al.}(2016)\citenamefont {Qian},
  \citenamefont {Xiao},\ and\ \citenamefont {Liu}}]{qian16}%
  \BibitemOpen
  \bibfield  {author} {\bibinfo {author} {\bibfnamefont {Haoliang}\
  \bibnamefont {Qian}}, \bibinfo {author} {\bibfnamefont {Yuzhe}\ \bibnamefont
  {Xiao}}, \ and\ \bibinfo {author} {\bibfnamefont {Zhaowei}\ \bibnamefont
  {Liu}},\ }\bibfield  {title} {\enquote {\bibinfo {title} {Giant {Kerr}
  response of ultrathin gold films from quantum size effect},}\ }\href@noop {}
  {\bibfield  {journal} {\bibinfo  {journal} {Nat. Commun.}\ }\textbf {\bibinfo
  {volume} {7}},\ \bibinfo {pages} {1--6} (\bibinfo {year} {2016})}\BibitemShut
  {NoStop}%
\bibitem [{\citenamefont {Varas}\ \emph {et~al.}(2016)\citenamefont {Varas},
  \citenamefont {Garcia-Gonzalez}, \citenamefont {Feist}, \citenamefont
  {G.-Vidal},\ and\ \citenamefont {Rubio}}]{varas}%
  \BibitemOpen
  \bibfield  {author} {\bibinfo {author} {\bibfnamefont {A.}~\bibnamefont
  {Varas}}, \bibinfo {author} {\bibfnamefont {P.}~\bibnamefont
  {Garcia-Gonzalez}}, \bibinfo {author} {\bibfnamefont {J.}~\bibnamefont
  {Feist}}, \bibinfo {author} {\bibfnamefont {F.~J.}\ \bibnamefont {G.-Vidal}},
  \ and\ \bibinfo {author} {\bibfnamefont {A.}~\bibnamefont {Rubio}},\
  }\bibfield  {title} {\enquote {\bibinfo {title} {Quantum plasmonics: from
  jellium models to ab initio calculations},}\ }\href@noop {} {\bibfield
  {journal} {\bibinfo  {journal} {Nanophotonics}\ }\textbf {\bibinfo {volume}
  {5}},\ \bibinfo {pages} {409--426} (\bibinfo {year} {2016})}\BibitemShut
  {NoStop}%
\bibitem [{\citenamefont {Zhou}\ \emph {et~al.}(2021)\citenamefont {Zhou},
  \citenamefont {Du}, \citenamefont {Wang},\ and\ \citenamefont
  {Jin}}]{zhou21}%
  \BibitemOpen
  \bibfield  {author} {\bibinfo {author} {\bibfnamefont {Meng}\ \bibnamefont
  {Zhou}}, \bibinfo {author} {\bibfnamefont {Xiangsha}\ \bibnamefont {Du}},
  \bibinfo {author} {\bibfnamefont {He}~\bibnamefont {Wang}}, \ and\ \bibinfo
  {author} {\bibfnamefont {Rongchao}\ \bibnamefont {Jin}},\ }\bibfield  {title}
  {\enquote {\bibinfo {title} {The critical number of gold atoms for a metallic
  state nanocluster: Resolving a decades-long question},}\ }\href@noop {}
  {\bibfield  {journal} {\bibinfo  {journal} {{ACS} nano}\ }\textbf {\bibinfo
  {volume} {15}},\ \bibinfo {pages} {13980--13992} (\bibinfo {year}
  {2021})}\BibitemShut {NoStop}%
\bibitem [{\citenamefont {Wood}\ and\ \citenamefont {Ashcroft}(1982)}]{wood82}%
  \BibitemOpen
  \bibfield  {author} {\bibinfo {author} {\bibfnamefont {D.~M.}\ \bibnamefont
  {Wood}}\ and\ \bibinfo {author} {\bibfnamefont {N.~W.}\ \bibnamefont
  {Ashcroft}},\ }\bibfield  {title} {\enquote {\bibinfo {title} {Quantum size
  effects in the optical properties of small metallic particles},}\ }\href
  {\doibase 10.1103/PhysRevB.25.6255} {\bibfield  {journal} {\bibinfo
  {journal} {Phys. Rev. B}\ }\textbf {\bibinfo {volume} {25}},\ \bibinfo
  {pages} {6255--6274} (\bibinfo {year} {1982})}\BibitemShut {NoStop}%
\bibitem [{\citenamefont {Hache}\ \emph {et~al.}(1986)\citenamefont {Hache},
  \citenamefont {Ricard},\ and\ \citenamefont {Flytzanis}}]{hache86}%
  \BibitemOpen
  \bibfield  {author} {\bibinfo {author} {\bibfnamefont {F.}~\bibnamefont
  {Hache}}, \bibinfo {author} {\bibfnamefont {D.}~\bibnamefont {Ricard}}, \
  and\ \bibinfo {author} {\bibfnamefont {Ch.}\ \bibnamefont {Flytzanis}},\
  }\bibfield  {title} {\enquote {\bibinfo {title} {Optical nonlinearities of
  small metal particles: surface-mediated resonance and quantum size
  effects},}\ }\href@noop {} {\bibfield  {journal} {\bibinfo  {journal} {JOSA
  B}\ }\textbf {\bibinfo {volume} {3}},\ \bibinfo {pages} {1647--1655}
  (\bibinfo {year} {1986})}\BibitemShut {NoStop}%
\bibitem [{\citenamefont {Sato}\ \emph {et~al.}(2015)\citenamefont {Sato},
  \citenamefont {Ohnuma}, \citenamefont {Oyoshi},\ and\ \citenamefont
  {Takeda}}]{sato}%
  \BibitemOpen
  \bibfield  {author} {\bibinfo {author} {\bibfnamefont {Rodrigo}\ \bibnamefont
  {Sato}}, \bibinfo {author} {\bibfnamefont {Masato}\ \bibnamefont {Ohnuma}},
  \bibinfo {author} {\bibfnamefont {Keiji}\ \bibnamefont {Oyoshi}}, \ and\
  \bibinfo {author} {\bibfnamefont {Yoshihiko}\ \bibnamefont {Takeda}},\
  }\bibfield  {title} {\enquote {\bibinfo {title} {Spectral investigation of
  nonlinear local field effects in {Ag} nanoparticles},}\ }\href {\doibase
  10.1063/1.4914907} {\bibfield  {journal} {\bibinfo  {journal} {J. Appl.
  Phys.}\ }\textbf {\bibinfo {volume} {117}},\ \bibinfo {pages} {113101}
  (\bibinfo {year} {2015})}\BibitemShut {NoStop}%
\bibitem [{\citenamefont {Genzel}\ \emph {et~al.}(1975)\citenamefont {Genzel},
  \citenamefont {Martin},\ and\ \citenamefont {Kreibig}}]{genzel75}%
  \BibitemOpen
  \bibfield  {author} {\bibinfo {author} {\bibfnamefont {L.}~\bibnamefont
  {Genzel}}, \bibinfo {author} {\bibfnamefont {T.~P.}\ \bibnamefont {Martin}},
  \ and\ \bibinfo {author} {\bibfnamefont {U.}~\bibnamefont {Kreibig}},\
  }\bibfield  {title} {\enquote {\bibinfo {title} {Dielectric function and
  plasma resonances of small metal particles},}\ }\href@noop {} {\bibfield
  {journal} {\bibinfo  {journal} {Zeitschrift f{\"u}r Physik B Condensed
  Matter}\ }\textbf {\bibinfo {volume} {21}},\ \bibinfo {pages} {339--346}
  (\bibinfo {year} {1975})}\BibitemShut {NoStop}%
\bibitem [{\citenamefont {Brack}(1993)}]{brack93}%
  \BibitemOpen
  \bibfield  {author} {\bibinfo {author} {\bibfnamefont {Matthias}\
  \bibnamefont {Brack}},\ }\bibfield  {title} {\enquote {\bibinfo {title} {The
  physics of simple metal clusters: self-consistent jellium model and
  semiclassical approaches},}\ }\href {\doibase 10.1103/RevModPhys.65.677}
  {\bibfield  {journal} {\bibinfo  {journal} {Rev. Mod. Phys.}\ }\textbf
  {\bibinfo {volume} {65}},\ \bibinfo {pages} {677--732} (\bibinfo {year}
  {1993})}\BibitemShut {NoStop}%
\bibitem [{\citenamefont {Ginzburg}\ \emph {et~al.}(2014)\citenamefont
  {Ginzburg}, \citenamefont {Krasavin}, \citenamefont {Wurtz},\ and\
  \citenamefont {Zayats}}]{ginzburg14-hydrodynamics-nanostr}%
  \BibitemOpen
  \bibfield  {author} {\bibinfo {author} {\bibfnamefont {Pavel}\ \bibnamefont
  {Ginzburg}}, \bibinfo {author} {\bibfnamefont {Alexey~V}\ \bibnamefont
  {Krasavin}}, \bibinfo {author} {\bibfnamefont {Gregory~A}\ \bibnamefont
  {Wurtz}}, \ and\ \bibinfo {author} {\bibfnamefont {Anatoly~V}\ \bibnamefont
  {Zayats}},\ }\bibfield  {title} {\enquote {\bibinfo {title} {Nonperturbative
  hydrodynamic model for multiple harmonics generation in metallic
  nanostructures},}\ }\href@noop {} {\bibfield  {journal} {\bibinfo  {journal}
  {{ACS} Photonics}\ }\textbf {\bibinfo {volume} {2}},\ \bibinfo {pages}
  {8--13} (\bibinfo {year} {2014})}\BibitemShut {NoStop}%
\bibitem [{\citenamefont {Hurst}\ \emph {et~al.}(2014)\citenamefont {Hurst},
  \citenamefont {Haas}, \citenamefont {Manfredi},\ and\ \citenamefont
  {Hervieux}}]{hurst14-hydrodyn-nanostr}%
  \BibitemOpen
  \bibfield  {author} {\bibinfo {author} {\bibfnamefont {J\'er\^ome}\
  \bibnamefont {Hurst}}, \bibinfo {author} {\bibfnamefont {Fernando}\
  \bibnamefont {Haas}}, \bibinfo {author} {\bibfnamefont {Giovanni}\
  \bibnamefont {Manfredi}}, \ and\ \bibinfo {author} {\bibfnamefont
  {Paul-Antoine}\ \bibnamefont {Hervieux}},\ }\bibfield  {title} {\enquote
  {\bibinfo {title} {High-harmonic generation by nonlinear resonant excitation
  of surface plasmon modes in metallic nanoparticles},}\ }\href {\doibase
  10.1103/PhysRevB.89.161111} {\bibfield  {journal} {\bibinfo  {journal} {Phys.
  Rev. B}\ }\textbf {\bibinfo {volume} {89}},\ \bibinfo {pages} {161111}
  (\bibinfo {year} {2014})}\BibitemShut {NoStop}%
\bibitem [{\citenamefont {Takeuchi}\ and\ \citenamefont
  {Yabana}(2022)}]{Takeuchi}%
  \BibitemOpen
  \bibfield  {author} {\bibinfo {author} {\bibfnamefont {Takashi}\ \bibnamefont
  {Takeuchi}}\ and\ \bibinfo {author} {\bibfnamefont {Kazuhiro}\ \bibnamefont
  {Yabana}},\ }\bibfield  {title} {\enquote {\bibinfo {title} {Electron
  spill-out effect on third-order optical nonlinearity of metallic
  nanostructure},}\ }\href {\doibase 10.1103/PhysRevA.106.063517} {\bibfield
  {journal} {\bibinfo  {journal} {Phys. Rev. A}\ }\textbf {\bibinfo {volume}
  {106}},\ \bibinfo {pages} {063517} (\bibinfo {year} {2022})}\BibitemShut
  {NoStop}%
\bibitem [{\citenamefont {Zhang}\ \emph {et~al.}(2017)\citenamefont {Zhang},
  \citenamefont {Xiang}, \citenamefont {Zhang},\ and\ \citenamefont
  {Lu}}]{zhang17}%
  \BibitemOpen
  \bibfield  {author} {\bibinfo {author} {\bibfnamefont {Xu}~\bibnamefont
  {Zhang}}, \bibinfo {author} {\bibfnamefont {Hongping}\ \bibnamefont {Xiang}},
  \bibinfo {author} {\bibfnamefont {Mingliang}\ \bibnamefont {Zhang}}, \ and\
  \bibinfo {author} {\bibfnamefont {Gang}\ \bibnamefont {Lu}},\ }\bibfield
  {title} {\enquote {\bibinfo {title} {Plasmonic resonances of nanoparticles
  from large-scale quantum mechanical simulations},}\ }\href@noop {} {\bibfield
   {journal} {\bibinfo  {journal} {International Journal of Modern Physics B}\
  }\textbf {\bibinfo {volume} {31}},\ \bibinfo {pages} {1740003} (\bibinfo
  {year} {2017})}\BibitemShut {NoStop}%
\bibitem [{\citenamefont {Barbry}\ \emph {et~al.}(2015)\citenamefont {Barbry},
  \citenamefont {Koval}, \citenamefont {Marchesin}, \citenamefont {Esteban},
  \citenamefont {Borisov}, \citenamefont {Aizpurua},\ and\ \citenamefont
  {Sanchez-Portal}}]{barbry}%
  \BibitemOpen
  \bibfield  {author} {\bibinfo {author} {\bibfnamefont {M.}~\bibnamefont
  {Barbry}}, \bibinfo {author} {\bibfnamefont {P.}~\bibnamefont {Koval}},
  \bibinfo {author} {\bibfnamefont {F}~\bibnamefont {Marchesin}}, \bibinfo
  {author} {\bibfnamefont {R.}~\bibnamefont {Esteban}}, \bibinfo {author}
  {\bibfnamefont {A.~G.}\ \bibnamefont {Borisov}}, \bibinfo {author}
  {\bibfnamefont {J.}~\bibnamefont {Aizpurua}}, \ and\ \bibinfo {author}
  {\bibfnamefont {D}~\bibnamefont {Sanchez-Portal}},\ }\bibfield  {title}
  {\enquote {\bibinfo {title} {Atomistic near-field nanoplasmonics: Reaching
  atomic-scale resolution in nanooptics},}\ }\href@noop {} {\bibfield
  {journal} {\bibinfo  {journal} {Nano Lett.}\ }\textbf {\bibinfo {volume}
  {15}},\ \bibinfo {pages} {3410--3419} (\bibinfo {year} {2015})}\BibitemShut
  {NoStop}%
\bibitem [{\citenamefont {Rossi}\ \emph {et~al.}(2017)\citenamefont {Rossi},
  \citenamefont {Kuisma}, \citenamefont {Puska}, \citenamefont {M.},\ and\
  \citenamefont {Erhart}}]{rossi}%
  \BibitemOpen
  \bibfield  {author} {\bibinfo {author} {\bibfnamefont {T.~P.}\ \bibnamefont
  {Rossi}}, \bibinfo {author} {\bibfnamefont {M}~\bibnamefont {Kuisma}},
  \bibinfo {author} {\bibfnamefont {M.~J.}\ \bibnamefont {Puska}}, \bibinfo
  {author} {\bibfnamefont {Nieminen~R.}\ \bibnamefont {M.}}, \ and\ \bibinfo
  {author} {\bibfnamefont {P.}~\bibnamefont {Erhart}},\ }\bibfield  {title}
  {\enquote {\bibinfo {title} {Kohn–sham decomposition in real-time
  time-dependent density-functional theory: An efficient tool for analyzing
  plasmonic excitation},}\ }\href@noop {} {\bibfield  {journal} {\bibinfo
  {journal} {J. Chem. Theory Comput.}\ }\textbf {\bibinfo {volume} {13}},\
  \bibinfo {pages} {4779–4790} (\bibinfo {year} {2017})}\BibitemShut
  {NoStop}%
\bibitem [{\citenamefont {Day}\ \emph {et~al.}(2010)\citenamefont {Day},
  \citenamefont {Nguyen},\ and\ \citenamefont {Pachter}}]{day10}%
  \BibitemOpen
  \bibfield  {author} {\bibinfo {author} {\bibfnamefont {Paul~N}\ \bibnamefont
  {Day}}, \bibinfo {author} {\bibfnamefont {Kiet~A}\ \bibnamefont {Nguyen}}, \
  and\ \bibinfo {author} {\bibfnamefont {Ruth}\ \bibnamefont {Pachter}},\
  }\bibfield  {title} {\enquote {\bibinfo {title} {Calculation of one-and
  two-photon absorption spectra of thiolated gold nanoclusters using
  time-dependent density functional theory},}\ }\href@noop {} {\bibfield
  {journal} {\bibinfo  {journal} {J. Chem. Theor. Comput.}\ }\textbf {\bibinfo
  {volume} {6}},\ \bibinfo {pages} {2809--2821} (\bibinfo {year}
  {2010})}\BibitemShut {NoStop}%
\bibitem [{\citenamefont {Day}\ \emph {et~al.}(2016)\citenamefont {Day},
  \citenamefont {Pachter}, \citenamefont {Nguyen},\ and\ \citenamefont
  {Bigioni}}]{day16}%
  \BibitemOpen
  \bibfield  {author} {\bibinfo {author} {\bibfnamefont {Paul~N}\ \bibnamefont
  {Day}}, \bibinfo {author} {\bibfnamefont {Ruth}\ \bibnamefont {Pachter}},
  \bibinfo {author} {\bibfnamefont {Kiet~A}\ \bibnamefont {Nguyen}}, \ and\
  \bibinfo {author} {\bibfnamefont {Terry~P}\ \bibnamefont {Bigioni}},\
  }\bibfield  {title} {\enquote {\bibinfo {title} {Linear and nonlinear optical
  response in silver nanoclusters: insight from a computational
  investigation},}\ }\href@noop {} {\bibfield  {journal} {\bibinfo  {journal}
  {J. Phys. Chem. A}\ }\textbf {\bibinfo {volume} {120}},\ \bibinfo {pages}
  {507--518} (\bibinfo {year} {2016})}\BibitemShut {NoStop}%
\bibitem [{\citenamefont {Zhou}\ \emph {et~al.}(2016)\citenamefont {Zhou},
  \citenamefont {Zeng}, \citenamefont {Chen}, \citenamefont {Zhao},
  \citenamefont {Sfeir}, \citenamefont {Zhu},\ and\ \citenamefont
  {Jin}}]{zhou16}%
  \BibitemOpen
  \bibfield  {author} {\bibinfo {author} {\bibfnamefont {Meng}\ \bibnamefont
  {Zhou}}, \bibinfo {author} {\bibfnamefont {Chenjie}\ \bibnamefont {Zeng}},
  \bibinfo {author} {\bibfnamefont {Yuxiang}\ \bibnamefont {Chen}}, \bibinfo
  {author} {\bibfnamefont {Shuo}\ \bibnamefont {Zhao}}, \bibinfo {author}
  {\bibfnamefont {Matthew~Y}\ \bibnamefont {Sfeir}}, \bibinfo {author}
  {\bibfnamefont {Manzhou}\ \bibnamefont {Zhu}}, \ and\ \bibinfo {author}
  {\bibfnamefont {Rongchao}\ \bibnamefont {Jin}},\ }\bibfield  {title}
  {\enquote {\bibinfo {title} {Evolution from the plasmon to exciton state in
  ligand-protected atomically precise gold nanoparticles},}\ }\href@noop {}
  {\bibfield  {journal} {\bibinfo  {journal} {Nat. Commun.}\ }\textbf {\bibinfo
  {volume} {7}},\ \bibinfo {pages} {1--7} (\bibinfo {year} {2016})}\BibitemShut
  {NoStop}%
\bibitem [{\citenamefont {Townsend}\ and\ \citenamefont
  {Bryant}(2012)}]{townsend}%
  \BibitemOpen
  \bibfield  {author} {\bibinfo {author} {\bibfnamefont {E.}~\bibnamefont
  {Townsend}}\ and\ \bibinfo {author} {\bibfnamefont {G.~W.}\ \bibnamefont
  {Bryant}},\ }\bibfield  {title} {\enquote {\bibinfo {title} {Plasmonic
  properties of metallic nanoparticles: The effects of size quantization},}\
  }\href@noop {} {\bibfield  {journal} {\bibinfo  {journal} {Nano Lett.}\
  }\textbf {\bibinfo {volume} {12}},\ \bibinfo {pages} {429–434} (\bibinfo
  {year} {2012})}\BibitemShut {NoStop}%
\bibitem [{\citenamefont {Barcaro}\ \emph {et~al.}(2006)\citenamefont
  {Barcaro}, \citenamefont {Fortunelli}, \citenamefont {Rossi}, \citenamefont
  {Nita},\ and\ \citenamefont {Ferrando}}]{barcaro06}%
  \BibitemOpen
  \bibfield  {author} {\bibinfo {author} {\bibfnamefont {Giovanni}\
  \bibnamefont {Barcaro}}, \bibinfo {author} {\bibfnamefont {Alessandro}\
  \bibnamefont {Fortunelli}}, \bibinfo {author} {\bibfnamefont {Giulia}\
  \bibnamefont {Rossi}}, \bibinfo {author} {\bibfnamefont {Florin}\
  \bibnamefont {Nita}}, \ and\ \bibinfo {author} {\bibfnamefont {Riccardo}\
  \bibnamefont {Ferrando}},\ }\bibfield  {title} {\enquote {\bibinfo {title}
  {Electronic and structural shell closure in {AgCu} and {AuCu}
  nanoclusters},}\ }\href@noop {} {\bibfield  {journal} {\bibinfo  {journal}
  {J. Phys. Chem. B}\ }\textbf {\bibinfo {volume} {110}},\ \bibinfo {pages}
  {23197--23203} (\bibinfo {year} {2006})}\BibitemShut {NoStop}%
\bibitem [{\citenamefont {Kwak}\ \emph {et~al.}(2017)\citenamefont {Kwak},
  \citenamefont {Thanthirige}, \citenamefont {Pyo}, \citenamefont {Lee},\ and\
  \citenamefont {Ramakrishna}}]{kwak17}%
  \BibitemOpen
  \bibfield  {author} {\bibinfo {author} {\bibfnamefont {Kyuju}\ \bibnamefont
  {Kwak}}, \bibinfo {author} {\bibfnamefont {Viraj~Dhanushka}\ \bibnamefont
  {Thanthirige}}, \bibinfo {author} {\bibfnamefont {Kyunglim}\ \bibnamefont
  {Pyo}}, \bibinfo {author} {\bibfnamefont {Dongil}\ \bibnamefont {Lee}}, \
  and\ \bibinfo {author} {\bibfnamefont {Guda}\ \bibnamefont {Ramakrishna}},\
  }\bibfield  {title} {\enquote {\bibinfo {title} {Energy gap law for exciton
  dynamics in gold cluster molecules},}\ }\href@noop {} {\bibfield  {journal}
  {\bibinfo  {journal} {J. Phys. Chem. Lett.}\ }\textbf {\bibinfo {volume}
  {8}},\ \bibinfo {pages} {4898--4905} (\bibinfo {year} {2017})}\BibitemShut
  {NoStop}%
\bibitem [{\citenamefont {Zhou}\ \emph {et~al.}(2019)\citenamefont {Zhou},
  \citenamefont {Higaki}, \citenamefont {Li}, \citenamefont {Zeng},
  \citenamefont {Li}, \citenamefont {Sfeir},\ and\ \citenamefont
  {Jin}}]{zhou19}%
  \BibitemOpen
  \bibfield  {author} {\bibinfo {author} {\bibfnamefont {Meng}\ \bibnamefont
  {Zhou}}, \bibinfo {author} {\bibfnamefont {Tatsuya}\ \bibnamefont {Higaki}},
  \bibinfo {author} {\bibfnamefont {Yingwei}\ \bibnamefont {Li}}, \bibinfo
  {author} {\bibfnamefont {Chenjie}\ \bibnamefont {Zeng}}, \bibinfo {author}
  {\bibfnamefont {Qi}~\bibnamefont {Li}}, \bibinfo {author} {\bibfnamefont
  {Matthew~Y.}\ \bibnamefont {Sfeir}}, \ and\ \bibinfo {author} {\bibfnamefont
  {Rongchao}\ \bibnamefont {Jin}},\ }\bibfield  {title} {\enquote {\bibinfo
  {title} {Three-stage evolution from nonscalable to scalable optical
  properties of thiolate-protected gold nanoclusters},}\ }\href@noop {}
  {\bibfield  {journal} {\bibinfo  {journal} {J. Am. Chem. Soc.}\ }\textbf
  {\bibinfo {volume} {141}},\ \bibinfo {pages} {19754--19764} (\bibinfo {year}
  {2019})}\BibitemShut {NoStop}%
\bibitem [{\citenamefont {Oates}\ and\ \citenamefont
  {M{\"u}cklich}(2005)}]{oates05}%
  \BibitemOpen
  \bibfield  {author} {\bibinfo {author} {\bibfnamefont {T.~W.~H.}\
  \bibnamefont {Oates}}\ and\ \bibinfo {author} {\bibfnamefont
  {A.}~\bibnamefont {M{\"u}cklich}},\ }\bibfield  {title} {\enquote {\bibinfo
  {title} {Evolution of plasmon resonances during plasma deposition of silver
  nanoparticles},}\ }\href@noop {} {\bibfield  {journal} {\bibinfo  {journal}
  {Nanotechnology}\ }\textbf {\bibinfo {volume} {16}},\ \bibinfo {pages} {2606}
  (\bibinfo {year} {2005})}\BibitemShut {NoStop}%
\bibitem [{\citenamefont {Wyrwas}\ \emph {et~al.}(2007)\citenamefont {Wyrwas},
  \citenamefont {Alvarez}, \citenamefont {Khoury}, \citenamefont {Price},
  \citenamefont {Schaaff},\ and\ \citenamefont {Whetten}}]{wyrwas07}%
  \BibitemOpen
  \bibfield  {author} {\bibinfo {author} {\bibfnamefont {R.~B.}\ \bibnamefont
  {Wyrwas}}, \bibinfo {author} {\bibfnamefont {M.~M.}\ \bibnamefont {Alvarez}},
  \bibinfo {author} {\bibfnamefont {J.~T.}\ \bibnamefont {Khoury}}, \bibinfo
  {author} {\bibfnamefont {R.~C.}\ \bibnamefont {Price}}, \bibinfo {author}
  {\bibfnamefont {T.~G.}\ \bibnamefont {Schaaff}}, \ and\ \bibinfo {author}
  {\bibfnamefont {R.~L.}\ \bibnamefont {Whetten}},\ }\bibfield  {title}
  {\enquote {\bibinfo {title} {The colours of nanometric gold},}\ }\href@noop
  {} {\bibfield  {journal} {\bibinfo  {journal} {Eur. Phys. J. D}\ }\textbf
  {\bibinfo {volume} {43}},\ \bibinfo {pages} {91--95} (\bibinfo {year}
  {2007})}\BibitemShut {NoStop}%
\bibitem [{\citenamefont {He}\ and\ \citenamefont {Zeng}(2010)}]{he10}%
  \BibitemOpen
  \bibfield  {author} {\bibinfo {author} {\bibfnamefont {Yi}~\bibnamefont
  {He}}\ and\ \bibinfo {author} {\bibfnamefont {Taofang}\ \bibnamefont
  {Zeng}},\ }\bibfield  {title} {\enquote {\bibinfo {title} {First-principles
  study and model of dielectric functions of silver nanoparticles},}\
  }\href@noop {} {\bibfield  {journal} {\bibinfo  {journal} {J. Phys. Chem. C}\
  }\textbf {\bibinfo {volume} {114}},\ \bibinfo {pages} {18023--18030}
  (\bibinfo {year} {2010})}\BibitemShut {NoStop}%
\bibitem [{\citenamefont {St\"ockmann}(1999)}]{stoeckmann99:book}%
  \BibitemOpen
  \bibfield  {author} {\bibinfo {author} {\bibfnamefont {Hans-J\"urgen}\
  \bibnamefont {St\"ockmann}},\ }\href@noop {} {\emph {\bibinfo {title}
  {Quantum Chaos: An Introduction}}}\ (\bibinfo  {publisher} {Cambridge
  University Press},\ \bibinfo {address} {New York, USA},\ \bibinfo {year}
  {1999})\BibitemShut {NoStop}%
\bibitem [{\citenamefont {Simons}\ and\ \citenamefont
  {Altshuler}(1993)}]{simons93}%
  \BibitemOpen
  \bibfield  {author} {\bibinfo {author} {\bibfnamefont {B.~D.}\ \bibnamefont
  {Simons}}\ and\ \bibinfo {author} {\bibfnamefont {B.~L.}\ \bibnamefont
  {Altshuler}},\ }\bibfield  {title} {\enquote {\bibinfo {title}
  {Universalities in the spectra of disordered and chaotic systems},}\ }\href
  {\doibase 10.1103/PhysRevB.48.5422} {\bibfield  {journal} {\bibinfo
  {journal} {Phys. Rev. B}\ }\textbf {\bibinfo {volume} {48}},\ \bibinfo
  {pages} {5422--5438} (\bibinfo {year} {1993})}\BibitemShut {NoStop}%
\bibitem [{\citenamefont {Ravnik}\ \emph {et~al.}(2021)\citenamefont {Ravnik},
  \citenamefont {Vaskivskyi}, \citenamefont {Vodeb}, \citenamefont
  {Aupi{\v{c}}}, \citenamefont {Vaskivskyi}, \citenamefont {Gole{\v{z}}},
  \citenamefont {Gerasimenko}, \citenamefont {Kabanov},\ and\ \citenamefont
  {Mihailovic}}]{ravnik21}%
  \BibitemOpen
  \bibfield  {author} {\bibinfo {author} {\bibfnamefont {Jan}\ \bibnamefont
  {Ravnik}}, \bibinfo {author} {\bibfnamefont {Yevhenii}\ \bibnamefont
  {Vaskivskyi}}, \bibinfo {author} {\bibfnamefont {Jaka}\ \bibnamefont
  {Vodeb}}, \bibinfo {author} {\bibfnamefont {Polona}\ \bibnamefont
  {Aupi{\v{c}}}}, \bibinfo {author} {\bibfnamefont {Igor}\ \bibnamefont
  {Vaskivskyi}}, \bibinfo {author} {\bibfnamefont {Denis}\ \bibnamefont
  {Gole{\v{z}}}}, \bibinfo {author} {\bibfnamefont {Yaroslav}\ \bibnamefont
  {Gerasimenko}}, \bibinfo {author} {\bibfnamefont {Viktor}\ \bibnamefont
  {Kabanov}}, \ and\ \bibinfo {author} {\bibfnamefont {Dragan}\ \bibnamefont
  {Mihailovic}},\ }\bibfield  {title} {\enquote {\bibinfo {title} {Quantum
  billiards with correlated electrons confined in triangular transition metal
  dichalcogenide monolayer nanostructures},}\ }\href@noop {} {\bibfield
  {journal} {\bibinfo  {journal} {Nat. Commun.}\ }\textbf {\bibinfo {volume}
  {12}},\ \bibinfo {pages} {1--8} (\bibinfo {year} {2021})}\BibitemShut
  {NoStop}%
\bibitem [{\citenamefont {Nakamura}\ and\ \citenamefont
  {Thomas}(1988)}]{nakamura88}%
  \BibitemOpen
  \bibfield  {author} {\bibinfo {author} {\bibfnamefont {K.}~\bibnamefont
  {Nakamura}}\ and\ \bibinfo {author} {\bibfnamefont {H.}~\bibnamefont
  {Thomas}},\ }\bibfield  {title} {\enquote {\bibinfo {title} {Quantum billiard
  in a magnetic field: Chaos and diamagnetism},}\ }\href {\doibase
  10.1103/PhysRevLett.61.247} {\bibfield  {journal} {\bibinfo  {journal} {Phys.
  Rev. Lett.}\ }\textbf {\bibinfo {volume} {61}},\ \bibinfo {pages} {247--250}
  (\bibinfo {year} {1988})}\BibitemShut {NoStop}%
\bibitem [{\citenamefont {Jalabert}\ \emph {et~al.}(1990)\citenamefont
  {Jalabert}, \citenamefont {Baranger},\ and\ \citenamefont
  {Stone}}]{jalabert90-chaos-quant-dots}%
  \BibitemOpen
  \bibfield  {author} {\bibinfo {author} {\bibfnamefont {Rodolfo~A.}\
  \bibnamefont {Jalabert}}, \bibinfo {author} {\bibfnamefont {Harold~U.}\
  \bibnamefont {Baranger}}, \ and\ \bibinfo {author} {\bibfnamefont
  {A.~Douglas}\ \bibnamefont {Stone}},\ }\bibfield  {title} {\enquote {\bibinfo
  {title} {Conductance fluctuations in the ballistic regime: A probe of quantum
  chaos?}}\ }\href {\doibase 10.1103/PhysRevLett.65.2442} {\bibfield  {journal}
  {\bibinfo  {journal} {Phys. Rev. Lett.}\ }\textbf {\bibinfo {volume} {65}},\
  \bibinfo {pages} {2442--2445} (\bibinfo {year} {1990})}\BibitemShut {NoStop}%
\bibitem [{\citenamefont {Akis}\ \emph {et~al.}(1997)\citenamefont {Akis},
  \citenamefont {Ferry},\ and\ \citenamefont {Bird}}]{akis97}%
  \BibitemOpen
  \bibfield  {author} {\bibinfo {author} {\bibfnamefont {R.}~\bibnamefont
  {Akis}}, \bibinfo {author} {\bibfnamefont {D.~K.}\ \bibnamefont {Ferry}}, \
  and\ \bibinfo {author} {\bibfnamefont {J.~P.}\ \bibnamefont {Bird}},\
  }\bibfield  {title} {\enquote {\bibinfo {title} {Wave function scarring
  effects in open stadium shaped quantum dots},}\ }\href {\doibase
  10.1103/PhysRevLett.79.123} {\bibfield  {journal} {\bibinfo  {journal} {Phys.
  Rev. Lett.}\ }\textbf {\bibinfo {volume} {79}},\ \bibinfo {pages} {123--126}
  (\bibinfo {year} {1997})}\BibitemShut {NoStop}%
\bibitem [{\citenamefont {Zozoulenko}\ and\ \citenamefont
  {Berggren}(1997)}]{zozoulenko97}%
  \BibitemOpen
  \bibfield  {author} {\bibinfo {author} {\bibfnamefont {I.~V.}\ \bibnamefont
  {Zozoulenko}}\ and\ \bibinfo {author} {\bibfnamefont {K.-F.}\ \bibnamefont
  {Berggren}},\ }\bibfield  {title} {\enquote {\bibinfo {title} {Quantum
  scattering, resonant states, and conductance fluctuations in an open square
  electron billiard},}\ }\href {\doibase 10.1103/PhysRevB.56.6931} {\bibfield
  {journal} {\bibinfo  {journal} {Phys. Rev. B}\ }\textbf {\bibinfo {volume}
  {56}},\ \bibinfo {pages} {6931--6941} (\bibinfo {year} {1997})}\BibitemShut
  {NoStop}%
\bibitem [{\citenamefont {Burke}\ \emph {et~al.}(2010)\citenamefont {Burke},
  \citenamefont {Akis}, \citenamefont {Day}, \citenamefont {Speyer},
  \citenamefont {Ferry},\ and\ \citenamefont
  {Bennett}}]{burke10-scars-quantum-dots-experiment}%
  \BibitemOpen
  \bibfield  {author} {\bibinfo {author} {\bibfnamefont {A.~M.}\ \bibnamefont
  {Burke}}, \bibinfo {author} {\bibfnamefont {R.}~\bibnamefont {Akis}},
  \bibinfo {author} {\bibfnamefont {T.~E.}\ \bibnamefont {Day}}, \bibinfo
  {author} {\bibfnamefont {Gil}\ \bibnamefont {Speyer}}, \bibinfo {author}
  {\bibfnamefont {D.~K.}\ \bibnamefont {Ferry}}, \ and\ \bibinfo {author}
  {\bibfnamefont {B.~R.}\ \bibnamefont {Bennett}},\ }\bibfield  {title}
  {\enquote {\bibinfo {title} {Periodic scarred states in open quantum dots as
  evidence of quantum darwinism},}\ }\href {\doibase
  10.1103/PhysRevLett.104.176801} {\bibfield  {journal} {\bibinfo  {journal}
  {Phys. Rev. Lett.}\ }\textbf {\bibinfo {volume} {104}},\ \bibinfo {pages}
  {176801} (\bibinfo {year} {2010})}\BibitemShut {NoStop}%
\bibitem [{\citenamefont {Ponomarenko}\ \emph {et~al.}(2008)\citenamefont
  {Ponomarenko}, \citenamefont {Schedin}, \citenamefont {Katsnelson},
  \citenamefont {Yang}, \citenamefont {Hill}, \citenamefont {Novoselov},\ and\
  \citenamefont {Geim}}]{ponomarenko08}%
  \BibitemOpen
  \bibfield  {author} {\bibinfo {author} {\bibfnamefont {L.~A.}\ \bibnamefont
  {Ponomarenko}}, \bibinfo {author} {\bibfnamefont {F.}~\bibnamefont
  {Schedin}}, \bibinfo {author} {\bibfnamefont {M.~I.}\ \bibnamefont
  {Katsnelson}}, \bibinfo {author} {\bibfnamefont {R.}~\bibnamefont {Yang}},
  \bibinfo {author} {\bibfnamefont {E.~W.}\ \bibnamefont {Hill}}, \bibinfo
  {author} {\bibfnamefont {K.~S.}\ \bibnamefont {Novoselov}}, \ and\ \bibinfo
  {author} {\bibfnamefont {A.~K.}\ \bibnamefont {Geim}},\ }\bibfield  {title}
  {\enquote {\bibinfo {title} {{Chaotic Dirac Billiard in Graphene Quantum
  Dots}},}\ }\href {\doibase 10.1126/science.1154663} {\bibfield  {journal}
  {\bibinfo  {journal} {Science}\ }\textbf {\bibinfo {volume} {320}},\ \bibinfo
  {pages} {356--358} (\bibinfo {year} {2008})}\BibitemShut {NoStop}%
\bibitem [{\citenamefont {Kucha{\v{r}}{\'\i}k}\ \emph
  {et~al.}(2019)\citenamefont {Kucha{\v{r}}{\'\i}k}, \citenamefont
  {N{\v{e}}mec},\ and\ \citenamefont {Ostatnick{\`y}}}]{kucharik19}%
  \BibitemOpen
  \bibfield  {author} {\bibinfo {author} {\bibfnamefont {Ji{\v{r}}{\'\i}}\
  \bibnamefont {Kucha{\v{r}}{\'\i}k}}, \bibinfo {author} {\bibfnamefont
  {Hynek}\ \bibnamefont {N{\v{e}}mec}}, \ and\ \bibinfo {author} {\bibfnamefont
  {Tom{\'a}{\v{s}}}\ \bibnamefont {Ostatnick{\`y}}},\ }\bibfield  {title}
  {\enquote {\bibinfo {title} {Terahertz conductivity and coupling between
  geometrical and plasmonic resonances in nanostructures},}\ }\href@noop {}
  {\bibfield  {journal} {\bibinfo  {journal} {Phys. Rev. B}\ }\textbf {\bibinfo
  {volume} {99}},\ \bibinfo {pages} {035407} (\bibinfo {year}
  {2019})}\BibitemShut {NoStop}%
\bibitem [{\citenamefont {Boyd}\ and\ \citenamefont
  {Prato}(2008)}]{boyd08:book}%
  \BibitemOpen
  \bibfield  {author} {\bibinfo {author} {\bibfnamefont {R.W.}\ \bibnamefont
  {Boyd}}\ and\ \bibinfo {author} {\bibfnamefont {D.}~\bibnamefont {Prato}},\
  }\href {https://books.google.de/books?id=uoRUi1Yb7ooC} {\emph {\bibinfo
  {title} {Nonlinear Optics}}}\ (\bibinfo  {publisher} {Elsevier Science},\
  \bibinfo {year} {2008})\BibitemShut {NoStop}%
\bibitem [{\citenamefont {Kabanov}\ and\ \citenamefont
  {Alexandrov}(2008)}]{kabanov08}%
  \BibitemOpen
  \bibfield  {author} {\bibinfo {author} {\bibfnamefont {V.~V.}\ \bibnamefont
  {Kabanov}}\ and\ \bibinfo {author} {\bibfnamefont {A.~S.}\ \bibnamefont
  {Alexandrov}},\ }\bibfield  {title} {\enquote {\bibinfo {title} {Electron
  relaxation in metals: Theory and exact analytical solutions},}\ }\href
  {\doibase 10.1103/PhysRevB.78.174514} {\bibfield  {journal} {\bibinfo
  {journal} {Phys. Rev. B}\ }\textbf {\bibinfo {volume} {78}},\ \bibinfo
  {pages} {174514} (\bibinfo {year} {2008})}\BibitemShut {NoStop}%
\bibitem [{\citenamefont {Kucha\v{r}\'{\i}k}\ and\ \citenamefont
  {N\v{e}mec}(2021)}]{Kucharik21}%
  \BibitemOpen
  \bibfield  {author} {\bibinfo {author} {\bibfnamefont {Ji\v{r}\'{\i}}\
  \bibnamefont {Kucha\v{r}\'{\i}k}}\ and\ \bibinfo {author} {\bibfnamefont
  {Hynek}\ \bibnamefont {N\v{e}mec}},\ }\bibfield  {title} {\enquote {\bibinfo
  {title} {Strong confinement-induced nonlinear terahertz response in
  semiconductor nanostructures revealed by monte carlo calculations},}\ }\href
  {\doibase 10.1103/PhysRevB.103.205426} {\bibfield  {journal} {\bibinfo
  {journal} {Phys. Rev. B}\ }\textbf {\bibinfo {volume} {103}},\ \bibinfo
  {pages} {205426} (\bibinfo {year} {2021})}\BibitemShut {NoStop}%
\bibitem [{\citenamefont {Zeng}\ \emph {et~al.}(1988)\citenamefont {Zeng},
  \citenamefont {Bergman}, \citenamefont {Hui},\ and\ \citenamefont
  {Stroud}}]{zeng88-effective-n-n2-nanoparticles}%
  \BibitemOpen
  \bibfield  {author} {\bibinfo {author} {\bibfnamefont {X.~C.}\ \bibnamefont
  {Zeng}}, \bibinfo {author} {\bibfnamefont {D.~J.}\ \bibnamefont {Bergman}},
  \bibinfo {author} {\bibfnamefont {P.~M.}\ \bibnamefont {Hui}}, \ and\
  \bibinfo {author} {\bibfnamefont {D.}~\bibnamefont {Stroud}},\ }\bibfield
  {title} {\enquote {\bibinfo {title} {Effective-medium theory for weakly
  nonlinear composites},}\ }\href@noop {} {\bibfield  {journal} {\bibinfo
  {journal} {Phys. Rev. B}\ }\textbf {\bibinfo {volume} {38}},\ \bibinfo
  {pages} {10970} (\bibinfo {year} {1988})}\BibitemShut {NoStop}%
\bibitem [{\citenamefont {Palik}(1998)}]{palik98:book}%
  \BibitemOpen
  \bibfield  {author} {\bibinfo {author} {\bibfnamefont {Edward~D}\
  \bibnamefont {Palik}},\ }\href@noop {} {\emph {\bibinfo {title} {Handbook of
  optical constants of solids}}},\ Vol.~\bibinfo {volume} {3}\ (\bibinfo
  {publisher} {Academic press},\ \bibinfo {year} {1998})\BibitemShut {NoStop}%
\bibitem [{\citenamefont {Poole}\ \emph {et~al.}(2013)\citenamefont {Poole},
  \citenamefont {Trendafilov}, \citenamefont {Shvets}, \citenamefont {Smith},\
  and\ \citenamefont {Chowdhury}}]{poole13-damage-gold}%
  \BibitemOpen
  \bibfield  {author} {\bibinfo {author} {\bibfnamefont {Patrick}\ \bibnamefont
  {Poole}}, \bibinfo {author} {\bibfnamefont {Simeon}\ \bibnamefont
  {Trendafilov}}, \bibinfo {author} {\bibfnamefont {Gennady}\ \bibnamefont
  {Shvets}}, \bibinfo {author} {\bibfnamefont {Douglas}\ \bibnamefont {Smith}},
  \ and\ \bibinfo {author} {\bibfnamefont {Enam}\ \bibnamefont {Chowdhury}},\
  }\bibfield  {title} {\enquote {\bibinfo {title} {Femtosecond laser damage
  threshold of pulse compression gratings for petawatt scale laser systems},}\
  }\href@noop {} {\bibfield  {journal} {\bibinfo  {journal} {Optics express}\
  }\textbf {\bibinfo {volume} {21}},\ \bibinfo {pages} {26341--26351} (\bibinfo
  {year} {2013})}\BibitemShut {NoStop}%
\bibitem [{\citenamefont {Chimier}\ \emph {et~al.}(2011)\citenamefont
  {Chimier}, \citenamefont {Ut{\'e}za}, \citenamefont {Sanner}, \citenamefont
  {Sentis}, \citenamefont {Itina}, \citenamefont {Lassonde}, \citenamefont
  {L{\'e}gar{\'e}}, \citenamefont {Vidal},\ and\ \citenamefont
  {Kieffer}}]{chimier11-damage-silica}%
  \BibitemOpen
  \bibfield  {author} {\bibinfo {author} {\bibfnamefont {B.}~\bibnamefont
  {Chimier}}, \bibinfo {author} {\bibfnamefont {O.}~\bibnamefont {Ut{\'e}za}},
  \bibinfo {author} {\bibfnamefont {N.}~\bibnamefont {Sanner}}, \bibinfo
  {author} {\bibfnamefont {M.}~\bibnamefont {Sentis}}, \bibinfo {author}
  {\bibfnamefont {T.}~\bibnamefont {Itina}}, \bibinfo {author} {\bibfnamefont
  {P.}~\bibnamefont {Lassonde}}, \bibinfo {author} {\bibfnamefont
  {F.}~\bibnamefont {L{\'e}gar{\'e}}}, \bibinfo {author} {\bibfnamefont
  {F.}~\bibnamefont {Vidal}}, \ and\ \bibinfo {author} {\bibfnamefont {J.~C.}\
  \bibnamefont {Kieffer}},\ }\bibfield  {title} {\enquote {\bibinfo {title}
  {Damage and ablation thresholds of fused-silica in femtosecond regime},}\
  }\href@noop {} {\bibfield  {journal} {\bibinfo  {journal} {Phys. Rev. B}\
  }\textbf {\bibinfo {volume} {84}},\ \bibinfo {pages} {094104} (\bibinfo
  {year} {2011})}\BibitemShut {NoStop}%
\bibitem [{\citenamefont {Luo}\ \emph {et~al.}(2014)\citenamefont {Luo},
  \citenamefont {Chatzakis}, \citenamefont {Wang}, \citenamefont {Niesler},
  \citenamefont {Wegener}, \citenamefont {Koschny},\ and\ \citenamefont
  {Soukoulis}}]{luo14}%
  \BibitemOpen
  \bibfield  {author} {\bibinfo {author} {\bibfnamefont {Liang}\ \bibnamefont
  {Luo}}, \bibinfo {author} {\bibfnamefont {Ioannis}\ \bibnamefont
  {Chatzakis}}, \bibinfo {author} {\bibfnamefont {Jigang}\ \bibnamefont
  {Wang}}, \bibinfo {author} {\bibfnamefont {Fabian~BP}\ \bibnamefont
  {Niesler}}, \bibinfo {author} {\bibfnamefont {Martin}\ \bibnamefont
  {Wegener}}, \bibinfo {author} {\bibfnamefont {Thomas}\ \bibnamefont
  {Koschny}}, \ and\ \bibinfo {author} {\bibfnamefont {Costas~M}\ \bibnamefont
  {Soukoulis}},\ }\bibfield  {title} {\enquote {\bibinfo {title} {Broadband
  terahertz generation from metamaterials},}\ }\href@noop {} {\bibfield
  {journal} {\bibinfo  {journal} {Nat. Commun.}\ }\textbf {\bibinfo {volume}
  {5}},\ \bibinfo {pages} {3055} (\bibinfo {year} {2014})}\BibitemShut
  {NoStop}%
\bibitem [{\citenamefont {Keren-Zur}\ \emph {et~al.}(2019)\citenamefont
  {Keren-Zur}, \citenamefont {Tal}, \citenamefont {Fleischer}, \citenamefont
  {Mittleman},\ and\ \citenamefont {Ellenbogen}}]{keren-zur19}%
  \BibitemOpen
  \bibfield  {author} {\bibinfo {author} {\bibfnamefont {Shay}\ \bibnamefont
  {Keren-Zur}}, \bibinfo {author} {\bibfnamefont {Mai}\ \bibnamefont {Tal}},
  \bibinfo {author} {\bibfnamefont {Sharly}\ \bibnamefont {Fleischer}},
  \bibinfo {author} {\bibfnamefont {Daniel~M}\ \bibnamefont {Mittleman}}, \
  and\ \bibinfo {author} {\bibfnamefont {Tal}\ \bibnamefont {Ellenbogen}},\
  }\bibfield  {title} {\enquote {\bibinfo {title} {Generation of
  spatiotemporally tailored terahertz wavepackets by nonlinear metasurfaces},}\
  }\href@noop {} {\bibfield  {journal} {\bibinfo  {journal} {Nat. Commun.}\
  }\textbf {\bibinfo {volume} {10}},\ \bibinfo {pages} {1778} (\bibinfo {year}
  {2019})}\BibitemShut {NoStop}%
\bibitem [{\citenamefont {Samizadeh~Nikoo}\ and\ \citenamefont
  {Matioli}(2023)}]{samizadeh23}%
  \BibitemOpen
  \bibfield  {author} {\bibinfo {author} {\bibfnamefont {Mohammad}\
  \bibnamefont {Samizadeh~Nikoo}}\ and\ \bibinfo {author} {\bibfnamefont
  {Elison}\ \bibnamefont {Matioli}},\ }\bibfield  {title} {\enquote {\bibinfo
  {title} {Electronic metadevices for terahertz applications},}\ }\href@noop {}
  {\bibfield  {journal} {\bibinfo  {journal} {Nature}\ }\textbf {\bibinfo
  {volume} {614}},\ \bibinfo {pages} {451--455} (\bibinfo {year}
  {2023})}\BibitemShut {NoStop}%
\bibitem [{\citenamefont {Okamoto}(2021)}]{okamoto21:book}%
  \BibitemOpen
  \bibfield  {author} {\bibinfo {author} {\bibfnamefont {Katsunari}\
  \bibnamefont {Okamoto}},\ }\href@noop {} {\emph {\bibinfo {title}
  {Fundamentals of optical waveguides}}}\ (\bibinfo  {publisher} {Elsevier},\
  \bibinfo {year} {2021})\BibitemShut {NoStop}%
\bibitem [{\citenamefont {Kitamura}(2015)}]{kitamura15}%
  \BibitemOpen
  \bibfield  {author} {\bibinfo {author} {\bibfnamefont {Hikaru}\ \bibnamefont
  {Kitamura}},\ }\bibfield  {title} {\enquote {\bibinfo {title} {Derivation of
  the drude conductivity from quantum kinetic equations},}\ }\href@noop {}
  {\bibfield  {journal} {\bibinfo  {journal} {Eur. J. Phys.}\ }\textbf
  {\bibinfo {volume} {36}},\ \bibinfo {pages} {065010} (\bibinfo {year}
  {2015})}\BibitemShut {NoStop}%
\bibitem [{\citenamefont {S{\"o}nnichsen}(2001)}]{sonnichsen2001plasmons:phd}%
  \BibitemOpen
  \bibfield  {author} {\bibinfo {author} {\bibfnamefont {Carsten}\ \bibnamefont
  {S{\"o}nnichsen}},\ }\emph {\bibinfo {title} {Plasmons in metal
  nanostructures}},\ \href@noop {} {Ph.D. thesis},\ \bibinfo  {school} {lmu}
  (\bibinfo {year} {2001})\BibitemShut {NoStop}%
\bibitem [{\citenamefont {Christensen}\ and\ \citenamefont
  {Seraphin}(1971)}]{christensen71}%
  \BibitemOpen
  \bibfield  {author} {\bibinfo {author} {\bibfnamefont {N.~Egede}\
  \bibnamefont {Christensen}}\ and\ \bibinfo {author} {\bibfnamefont {B.~O.}\
  \bibnamefont {Seraphin}},\ }\bibfield  {title} {\enquote {\bibinfo {title}
  {Relativistic band calculation and the optical properties of gold},}\ }\href
  {\doibase 10.1103/PhysRevB.4.3321} {\bibfield  {journal} {\bibinfo  {journal}
  {Phys. Rev. B}\ }\textbf {\bibinfo {volume} {4}},\ \bibinfo {pages}
  {3321--3344} (\bibinfo {year} {1971})}\BibitemShut {NoStop}%
\bibitem [{\citenamefont {Hache}\ \emph {et~al.}(1988)\citenamefont {Hache},
  \citenamefont {Ricard}, \citenamefont {Flytzanis},\ and\ \citenamefont
  {Kreibig}}]{hache88}%
  \BibitemOpen
  \bibfield  {author} {\bibinfo {author} {\bibfnamefont {F.}~\bibnamefont
  {Hache}}, \bibinfo {author} {\bibfnamefont {D}~\bibnamefont {Ricard}},
  \bibinfo {author} {\bibfnamefont {Ch}~\bibnamefont {Flytzanis}}, \ and\
  \bibinfo {author} {\bibfnamefont {U}~\bibnamefont {Kreibig}},\ }\bibfield
  {title} {\enquote {\bibinfo {title} {The optical {Kerr} effect in small metal
  particles and metal colloids: the case of gold},}\ }\href@noop {} {\bibfield
  {journal} {\bibinfo  {journal} {Appl. Phys. A}\ }\textbf {\bibinfo {volume}
  {47}},\ \bibinfo {pages} {347--357} (\bibinfo {year} {1988})}\BibitemShut
  {NoStop}%
\bibitem [{\citenamefont {Voisin}\ \emph {et~al.}(2000)\citenamefont {Voisin},
  \citenamefont {Christofilos}, \citenamefont {Del~Fatti}, \citenamefont
  {Vall\'ee}, \citenamefont {Pr\'evel}, \citenamefont {Cottancin},
  \citenamefont {Lerm\'e}, \citenamefont {Pellarin},\ and\ \citenamefont
  {Broyer}}]{voisin00}%
  \BibitemOpen
  \bibfield  {author} {\bibinfo {author} {\bibfnamefont {C.}~\bibnamefont
  {Voisin}}, \bibinfo {author} {\bibfnamefont {D.}~\bibnamefont
  {Christofilos}}, \bibinfo {author} {\bibfnamefont {N.}~\bibnamefont
  {Del~Fatti}}, \bibinfo {author} {\bibfnamefont {F.}~\bibnamefont {Vall\'ee}},
  \bibinfo {author} {\bibfnamefont {B.}~\bibnamefont {Pr\'evel}}, \bibinfo
  {author} {\bibfnamefont {E.}~\bibnamefont {Cottancin}}, \bibinfo {author}
  {\bibfnamefont {J.}~\bibnamefont {Lerm\'e}}, \bibinfo {author} {\bibfnamefont
  {M.}~\bibnamefont {Pellarin}}, \ and\ \bibinfo {author} {\bibfnamefont
  {M.}~\bibnamefont {Broyer}},\ }\bibfield  {title} {\enquote {\bibinfo {title}
  {Size-dependent electron-electron interactions in metal nanoparticles},}\
  }\href {\doibase 10.1103/PhysRevLett.85.2200} {\bibfield  {journal} {\bibinfo
   {journal} {Phys. Rev. Lett.}\ }\textbf {\bibinfo {volume} {85}},\ \bibinfo
  {pages} {2200--2203} (\bibinfo {year} {2000})}\BibitemShut {NoStop}%
\bibitem [{\citenamefont {Besteiro}\ \emph {et~al.}(2019)\citenamefont
  {Besteiro}, \citenamefont {Yu}, \citenamefont {Wang}, \citenamefont
  {Holleitner}, \citenamefont {Hartland}, \citenamefont {Wiederrecht},\ and\
  \citenamefont {Govorov}}]{besteiro19}%
  \BibitemOpen
  \bibfield  {author} {\bibinfo {author} {\bibfnamefont {Lucas~V}\ \bibnamefont
  {Besteiro}}, \bibinfo {author} {\bibfnamefont {Peng}\ \bibnamefont {Yu}},
  \bibinfo {author} {\bibfnamefont {Zhiming}\ \bibnamefont {Wang}}, \bibinfo
  {author} {\bibfnamefont {Alexander~W}\ \bibnamefont {Holleitner}}, \bibinfo
  {author} {\bibfnamefont {Gregory~V}\ \bibnamefont {Hartland}}, \bibinfo
  {author} {\bibfnamefont {Gary~P}\ \bibnamefont {Wiederrecht}}, \ and\
  \bibinfo {author} {\bibfnamefont {Alexander~O}\ \bibnamefont {Govorov}},\
  }\bibfield  {title} {\enquote {\bibinfo {title} {The fast and the furious:
  Ultrafast hot electrons in plasmonic metastructures. size and structure
  matter},}\ }\href@noop {} {\bibfield  {journal} {\bibinfo  {journal} {Nano
  Today}\ }\textbf {\bibinfo {volume} {27}},\ \bibinfo {pages} {120--145}
  (\bibinfo {year} {2019})}\BibitemShut {NoStop}%
\bibitem [{\citenamefont {Hartland}\ \emph {et~al.}(2017)\citenamefont
  {Hartland}, \citenamefont {Besteiro}, \citenamefont {Johns},\ and\
  \citenamefont {Govorov}}]{hartland17}%
  \BibitemOpen
  \bibfield  {author} {\bibinfo {author} {\bibfnamefont {Gregory~V}\
  \bibnamefont {Hartland}}, \bibinfo {author} {\bibfnamefont {Lucas~V}\
  \bibnamefont {Besteiro}}, \bibinfo {author} {\bibfnamefont {Paul}\
  \bibnamefont {Johns}}, \ and\ \bibinfo {author} {\bibfnamefont {Alexander~O}\
  \bibnamefont {Govorov}},\ }\bibfield  {title} {\enquote {\bibinfo {title}
  {What’s so hot about electrons in metal nanoparticles?}}\ }\href@noop {}
  {\bibfield  {journal} {\bibinfo  {journal} {ACS Energy Lett.}\ }\textbf
  {\bibinfo {volume} {2}},\ \bibinfo {pages} {1641--1653} (\bibinfo {year}
  {2017})}\BibitemShut {NoStop}%
\bibitem [{\citenamefont {Sun}\ \emph {et~al.}(1994)\citenamefont {Sun},
  \citenamefont {Vall{\'e}e}, \citenamefont {Acioli}, \citenamefont {Ippen},\
  and\ \citenamefont {Fujimoto}}]{sun94}%
  \BibitemOpen
  \bibfield  {author} {\bibinfo {author} {\bibfnamefont {C.-K.}\ \bibnamefont
  {Sun}}, \bibinfo {author} {\bibfnamefont {F.}~\bibnamefont {Vall{\'e}e}},
  \bibinfo {author} {\bibfnamefont {L.~H.}\ \bibnamefont {Acioli}}, \bibinfo
  {author} {\bibfnamefont {E.~P.}\ \bibnamefont {Ippen}}, \ and\ \bibinfo
  {author} {\bibfnamefont {J.~G.}\ \bibnamefont {Fujimoto}},\ }\bibfield
  {title} {\enquote {\bibinfo {title} {Femtosecond-tunable measurement of
  electron thermalization in gold},}\ }\href@noop {} {\bibfield  {journal}
  {\bibinfo  {journal} {Phys. Rev. B}\ }\textbf {\bibinfo {volume} {50}},\
  \bibinfo {pages} {15337} (\bibinfo {year} {1994})}\BibitemShut {NoStop}%
\bibitem [{\citenamefont {Rotenberg}\ \emph {et~al.}(2007)\citenamefont
  {Rotenberg}, \citenamefont {Bristow}, \citenamefont {Pfeiffer}, \citenamefont
  {Betz},\ and\ \citenamefont {van Driel}}]{rotenberg07}%
  \BibitemOpen
  \bibfield  {author} {\bibinfo {author} {\bibfnamefont {Nir}\ \bibnamefont
  {Rotenberg}}, \bibinfo {author} {\bibfnamefont {A.~D.}\ \bibnamefont
  {Bristow}}, \bibinfo {author} {\bibfnamefont {Markus}\ \bibnamefont
  {Pfeiffer}}, \bibinfo {author} {\bibfnamefont {Markus}\ \bibnamefont {Betz}},
  \ and\ \bibinfo {author} {\bibfnamefont {H.~M.}\ \bibnamefont {van Driel}},\
  }\bibfield  {title} {\enquote {\bibinfo {title} {Nonlinear absorption in au
  films: Role of thermal effects},}\ }\href {\doibase
  10.1103/PhysRevB.75.155426} {\bibfield  {journal} {\bibinfo  {journal} {Phys.
  Rev. B}\ }\textbf {\bibinfo {volume} {75}},\ \bibinfo {pages} {155426}
  (\bibinfo {year} {2007})}\BibitemShut {NoStop}%
\bibitem [{\citenamefont {Hertel}\ \emph {et~al.}(1996)\citenamefont {Hertel},
  \citenamefont {Knoesel}, \citenamefont {Wolf},\ and\ \citenamefont
  {Ertl}}]{hertel96}%
  \BibitemOpen
  \bibfield  {author} {\bibinfo {author} {\bibfnamefont {T.}~\bibnamefont
  {Hertel}}, \bibinfo {author} {\bibfnamefont {E.}~\bibnamefont {Knoesel}},
  \bibinfo {author} {\bibfnamefont {M.}~\bibnamefont {Wolf}}, \ and\ \bibinfo
  {author} {\bibfnamefont {G.}~\bibnamefont {Ertl}},\ }\bibfield  {title}
  {\enquote {\bibinfo {title} {Ultrafast electron dynamics at {Cu(111)}:
  Response of an electron gas to optical excitation},}\ }\href {\doibase
  10.1103/PhysRevLett.76.535} {\bibfield  {journal} {\bibinfo  {journal} {Phys.
  Rev. Lett.}\ }\textbf {\bibinfo {volume} {76}},\ \bibinfo {pages} {535--538}
  (\bibinfo {year} {1996})}\BibitemShut {NoStop}%
\end{thebibliography}

%

\end{document}